\documentclass[openacc]{rstransa}

\usepackage{ragged2e}
\usepackage{amsmath, amssymb, bm}
\usepackage{graphicx}
\usepackage{tablefootnote}
\usepackage{amsfonts}
\usepackage{lmodern}
\usepackage{xcolor}
\usepackage[final, markup=underlined]{changes}

\newcommand\alf{Alfv\'en}

\newcommand{\kms}{km\,s$^{-1}$}
\jname{rsta}
\Journal{Phil. Trans. R. Soc}
\titlehead{Research}

\begin{document}
\title{Nonlinear steepening of a fast magnetoacoustic wave in the vicinity of a coronal magnetic null point}

\author{%%%% Author details
Yu Zhong$^{1}$, Valery M. Nakariakov$^{1,2}$, Mariana C\'ecere$^{3,4}$ and Andrea Costa$^{3}$}

%%%%%%%%% Insert author address here
\address{$^{1}$Centre for Fusion, Space and Astrophysics, Department of Physics, University of Warwick, Coventry CV4 7AL, UK\\
$^{2}$ Centro de Investigacion en Astronom\'ia, Universidad Bernardo O'Higgins, Avenida Viel 1497, Santiago, Chile\\
$^{3}$Instituto de Astronomía Te\'orica y Experimental, CONICET-UNC, C\'ordoba, Argentina\\
$^{4}$Observatorio Astronómico de C\'ordoba, UNC, C\'ordoba, Argentina}

%%%% Subject entries to be placed here %%%%
\subject{Solar system, Plasma physics, Wave motion}

%%%% Keyword entries to be placed here %%%%

%%%% Insert corresponding author and its email address}
\corres{Valery M. Nakariakov\\
\email{V.Nakariakov@warwick.ac.uk}}

%%%% Abstract text to be placed here %%%%%%%%%%%%
\begin{abstract} %maximum of 200 words
The interaction of a fast magnetoacoustic wave with a magnetic null point is studied in the context of the sympathetic flare phenomenon. Attention is paid to steepening the wave caused by the finite-amplitude effects in a non-uniform plasma environment. The null point is modelled by a potential magnetic configuration without a guiding field. The equilibrium plasma density and temperature are taken to be constant. 
The fast wave is excited by an impulsive point source outside the distance at which the local Alfv\'en and sound speeds are equal to each other. The incoming fast wave approaches the null point along the bisector of the magnetic configuration, i.e., across the local field. The fast-speed non-uniformity around the null point causes the refraction of the incident fast wave. However, the segment of the incoming wave, which approaches the null point is locally plane. The decrease in the fast speed towards the null point amplifies the nonlinear deformation of the incoming wave. Hence, the fast wave can become subject to nonlinear dissipation at a distance from the null point and not reach it.
\end{abstract}
%%%%%%%%%%%%%%%%%%%%%%%%%%%
\maketitle
%%%%%%%%%% Insert the texts which can accomdate on firstpage in the tag "fmtext" %%%%%

%\begin{fmtext}
\section{Introduction}
\label{sec:int}

Sympathetic solar flares are multiple flares occurring in different active regions of the Sun, often separated by large distances but appearing to be causally connected. Sympathetic flares on the Sun have been observed for several decades, e.g., \cite{2002ApJ...574..434M, 2015SoPh..290.2943S}, see also \cite{2025ApJS..278....9B, 2025A&A...694A..74G} for recent studies and references therein. Likewise, a chain of successive coronal mass ejections could occur if the second (\lq\lq daughter\rq\rq) eruption is induced by a preceding mother flare or eruption at a remote location, e.g., \cite{2007ApJ...664L.131Z, 2017SoPh..292...64L}. The phenomenon of sympathetic flares is rather rare, occurring in only 5\% of events, as shown by recent estimations \cite{2025A&A...694A..74G}. The mechanism for the initiation of the daughter flare or eruption by a mother flare or eruption is still under investigation, see, e.g., \cite{2009AdSpR..43..739S, 2025SoPh..300...82G, 2025A&A...694A..74G}.
Some scenarios require magnetic connectivity of the flare sites, e.g., \cite{2011ApJ...739L..63T, 2016ApJ...820...16J}.

As another possibility, it has been suggested that a fast magnetoacoustic wave excited by a mother flare can reach a magnetic null point in the epicentre of a possible daughter flare, and cause there a spike of an electric current density \cite{2006A&A...452..343N}. If the current density exceeds a certain threshold, various plasma microinstabilities can occur that cause anomalous resistivity, see, e.g., \cite{2006PhPl...13h2304B, 2022NatCo..13.2954G, 2025SSRv..221...20G}. In turn, anomalous resistivity is generally considered as an ignition agent of fast magnetic reconnection in solar flares, e.g., \cite{1994ApJ...436L.197Y, 2011LRSP....8....6S, 2024A&A...683A..95F}. A similar scenario but involving a slow magnetoacoustic wave has been considered too \cite{2006SoPh..238..313C}, in particular, as a mechanism for the progression of a flaring energy release along the neutral line in two-ribbon flares \cite{2011ApJ...730L..27N}. Furthermore, the interaction of an externally excited fast wave with a null point can cause quasi-periodic pulsations of the flaring emission, see, e.g., \cite{2017ApJ...844..149K, 2021SSRv..217...66Z}.

The interaction of magnetohydrodynamic (MHD) waves with a magnetic null point has been investigated in numerous studies. From the perspective of MHD wave dynamics, a magnetic null point is a region of non-uniformity in the characteristic MHD wave speeds. It has been established that fast magnetoacoustic waves experience refraction near the null, causing the wavefront to wrap around the null point \cite{2004A&A...420.1129M, 2011SSRv..158..205M}.
Furthermore, in the vicinity of the layer where local Alfv\'en and sound speeds are equal to each other, there occurs a mutual linear transformation of fast and slow magnetoacoustic wave modes, see, e.g., \cite{2006A&A...459..641M, 2016SoPh..291.3185A, 2017ApJ...837...94T, 2024A&A...681A..43Y}, directly detected in observations \cite{2024NatCo..15.2667K}.  In addition, the spatial non-uniformity of the fast-mode speed, specifically, its decrease towards the null, causes the front slopes of the wave to travel more slowly than the rear slopes, effectively shortening the wavelength. This increases spatial gradients of perturbed physical quantities, particularly resulting in an enhanced local electric current density. The accumulation of current density in an incoming fast wave has been clearly demonstrated in, e.g., \cite{2006A&A...459..641M, 2006A&A...452..343N}.
The refraction of the fast wave can lead to its focusing near the null point, and when combined with the decreasing wavelength, results in an increase in wave amplitude. This amplification intensifies nonlinear effects, such as wave steepening due to nonlinear cascade \cite{2009A&A...493..227M, 2011A&A...531A..63G, 2012SoPh..280..561A, 2013A&A...555A..86T, 2018A&A...611A..10S}. The main features of fast-wave interaction with a magnetic null, originally established in 2D geometry, have also been shown to occur in fully 3D models \cite{2008SoPh..251..563M}.

As stated in \cite{2011A&A...531A..63G}, the efficiency of seeding anomalous resistivity in the vicinity of a magnetic null point depends on the amplitude and wavelength of the incoming fast wave. In particular, it is crucial whether the shock forms near the null point or at a significant distance from it.
This can be illustrated with the following analogy: consider ocean surface waves approaching a sandy beach. A wave with large amplitude and relatively short wavelength may overturn and break far from the shoreline, causing little to no impact. In contrast, a wave of comparable or even lower amplitude but much longer wavelength, such as a tsunami, can reach the shoreline and propagate inland, causing severe damage.
In \cite{2011A&A...531A..63G}, the shock formation process was modelled for a cylindrically symmetric incoming wave front. 
However, a segment of the incoming wave front that approaches the null point along the bisector, i.e., across the equilibrium magnetic field between two magnetic separatrices, can be rather planar, see, e.g., \cite{2006A&A...459..641M, 2024NatCo..15.2667K}. In this study, our aim is to study nonlinear steepening of an incoming wave in this regime. We restrict our attention to 2D modelling, which allows us to reveal basic features of the wave evolution. Furthermore, a 2D model suits well some coronal magnetic configurations, such as a pseudo-streamer, e.g., \cite{2022A&A...662A.113S}.  

The paper is organised as follows. In Section~\ref{sec:mod}, we describe the model and governing equations.
In Section~\ref{sec:an} we make analytical estimations that highlight the main features of the process of interest.
In Section~\ref{sec:num}, the results of numerical simulations are presented. 
The results obtained are summarised and discussed in Section~\ref{sec:con}. 

\section{Model and governing equations}
\label{sec:mod}

\begin{figure}
    \centering
    \includegraphics[width=0.51\linewidth,trim={0 0 0 0}, clip]{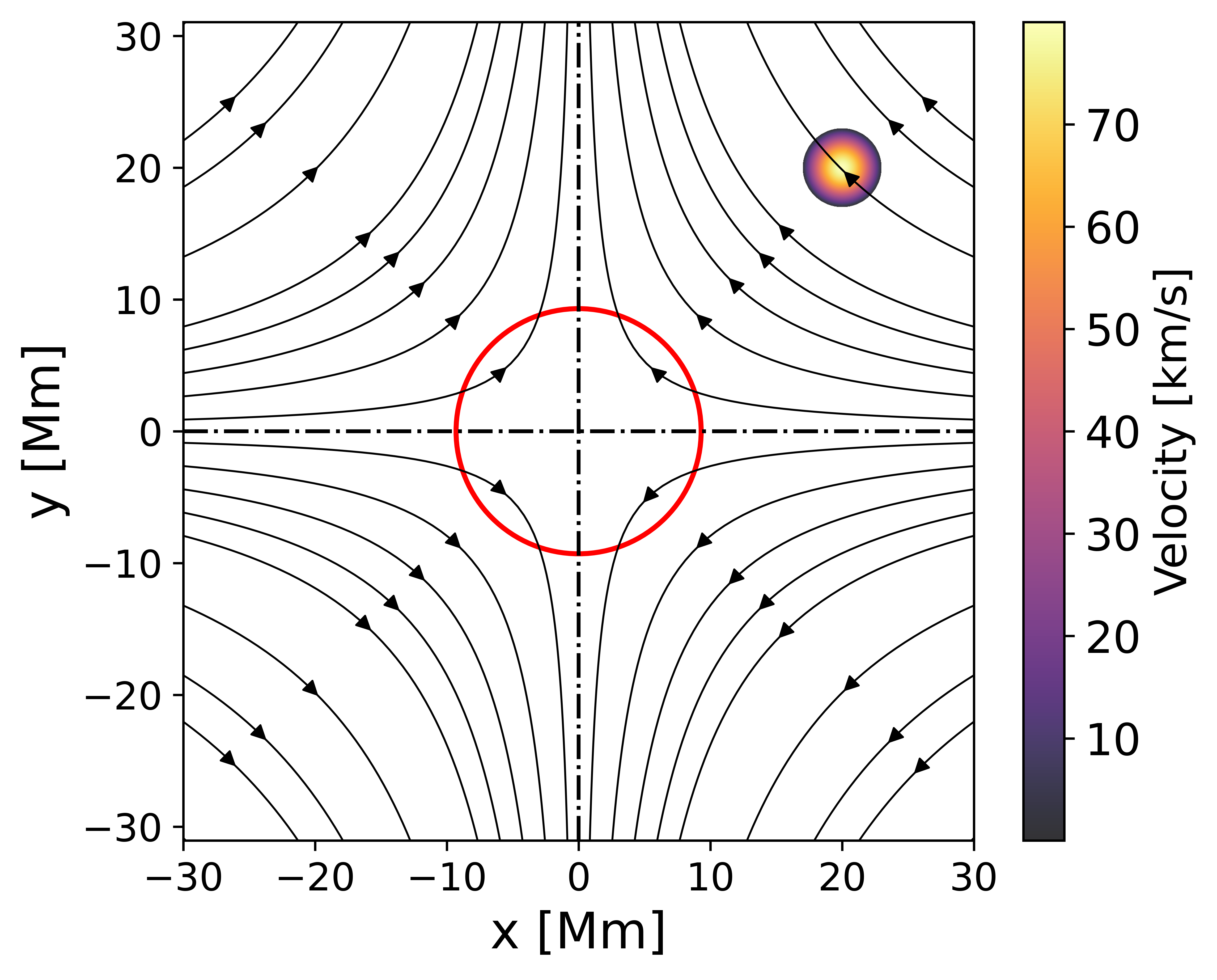}
    \includegraphics[width=0.465\linewidth,trim={0 0 0 0}, clip]{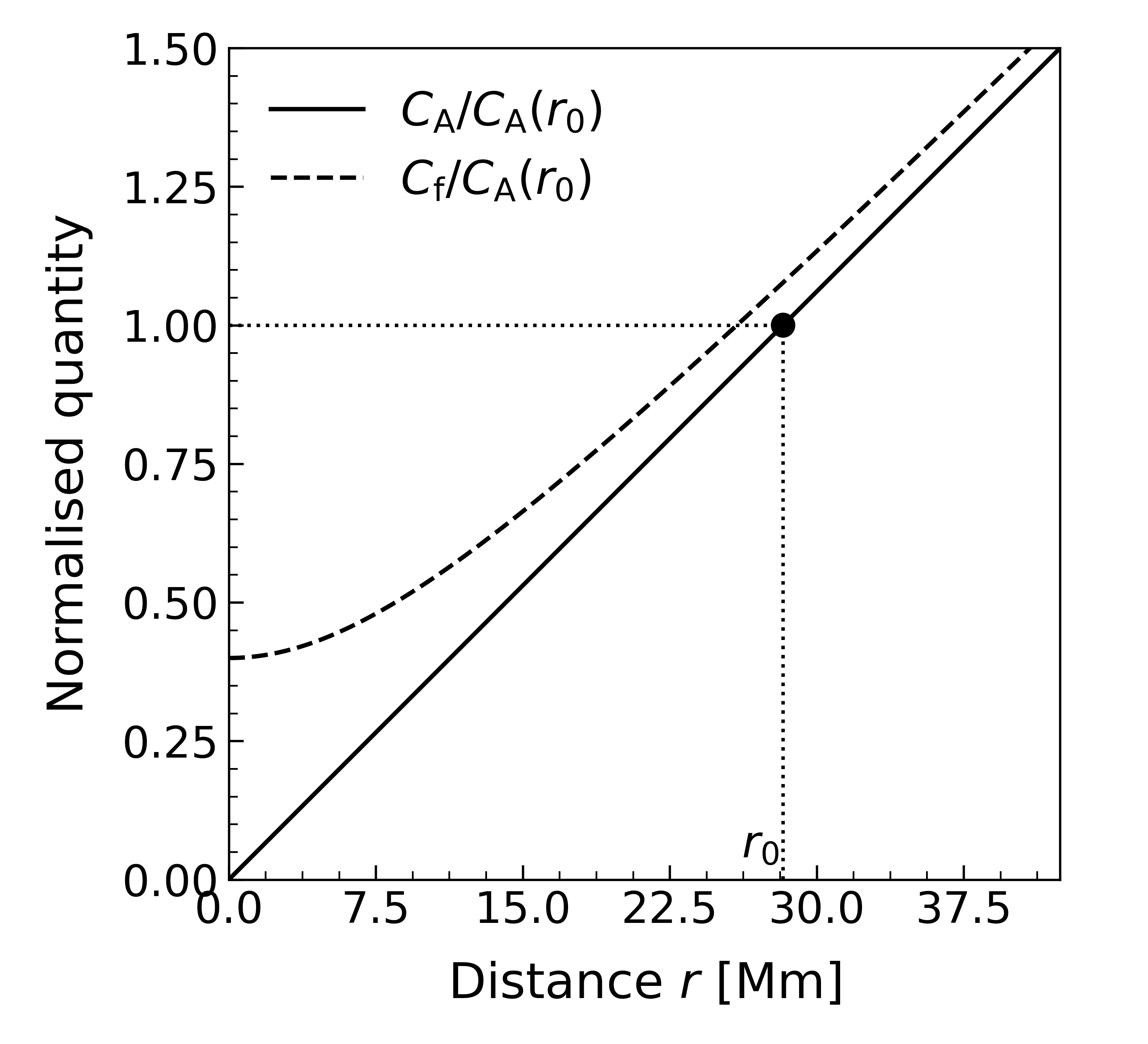}
    \caption{The equilibrium null-point magnetic field configuration (\textit{left panel}). The magnetic separatrices (dash-dot lines) and the $\beta \approx 1$ equipartition layer (red circle) are demonstrated. A shaded circular region outside the null point indicates the location of the initial pulse. The \alf \ speed ($C_\mathrm{A}$) and fast wave speed ($C_\mathrm{f}$) dependence on the distance ($r$) from the null-point (\textit{right panel}).}
    \label{fig:xp}
\end{figure}

Consider a 2D null point without the guiding field. 
The magnetic field configuration with a null point has a standard potential field geometry,
 \begin{equation}
 \label{Eq:beq}
 \pmb B  = \frac{B_0}{L_0}\displaystyle \Big(-x, y, 0 \Big),
 \end{equation}
where $B_0$ is the characteristic magnetic field strength, $L_0$ is the length scale of the field non-uniformity, and $x$ and $y$ are Cartesian axes with the origin at the null point, see Figure~\ref{fig:xp}. The field strength $B$ is zero at the null point and increases outward from the null point. 
For simplicity, the equilibrium density $\rho_0$ and temperature $T_0$ are taken to be constant. Thus, the sound speed $C_\mathrm{s}$ is constant, while the \alf~speed $C_\mathrm{A}$ is zero at the null point and increases radially outwards it. At a distance from the null point, the plasma parameter $\beta$ defined as the ratio of the sound and Alfv\'en speeds squared is much less than unity. This definition of the parameter $\beta$ is slightly different from the standard one, in which it is the ratio of the plasma thermal pressure to magnetic pressure. At the distance 
\begin{equation}
\label{req}
    r_\mathrm{eq} \approx \displaystyle \frac{L_0}{B_0} \Big( \frac{4 \pi k_\mathrm{B}}{\mu m_p}\rho_0 T_0\Big)^{1/2}, 
\end{equation}
where $k_\mathrm{B}$ is the Boltzmann constant, $\mu$ is mean molecular weight, $m_\mathrm{p}$ is the proton mass, and other quantities are in CGS units, the parameter $\beta$ is about unity, and the Alfv\'en and sound speeds are approximately equal to each other. Close to the null point, the plasma beta increases, approaching infinity. 

MHD perturbations of the equilibrium are described by the set of ideal MHD equations. In conservative form, the governing equations are as follows,
%\begin{displaymath}
\begin{align}
 %conservative form
\displaystyle  
& \frac{\partial \rho}{\partial t} + \nabla \cdot (\rho \pmb{v}) = 0,\\
& \frac{\partial (\rho \pmb{v})}{\partial t} + \nabla \cdot \Bigl(\rho \pmb{v} \pmb{v} + \bigl(p + \frac{|\pmb{B}|^2}{8\pi}\bigr)\,\pmb{I} - \frac{\pmb{B} \pmb{B}}{4\pi}\Bigr ) = 0,\\ 
& \frac{\partial \pmb{B}} {\partial t} + \nabla \cdot \bigl( \pmb{v} \pmb{B} - \pmb{B}  \pmb{v} \bigr) = 0,\\
& \frac{\partial \mathcal{E}}{\partial t} + \nabla\cdot\Bigl[\bigl(\mathcal{E} + p + \frac{|\pmb{B}|^2}{8\pi}\bigr)\pmb{v} -\frac{(\pmb{v} \cdot \pmb{B})  \pmb{B}} {4\pi} \Bigr] = 0,\\
&\nabla \cdot \pmb{B}=0,
\end{align}
%\end{displaymath}\cite{}
where the notations are standard, and $\mathcal{E}$ represents the total energy per unit volume. The thermodynamic quantities $\rho$, $p$ and $T$ are linked to each other via the ideal gas law. 

\section{Plane wave approximation}
\label{sec:an}

\subsection{Linear regime}
\label{sec:anlin}

As the segment of the fast wave front, which approaches the null point across the magnetic field is almost planar, its evolution can be approximately described by a 1D model. The equilibrium density $\rho_0$, pressure $p_0$ and temperature $T_0$, and hence the sound speed $C_\mathrm{s}$ are constant. 
Consider a fast wave propagating strictly across a non-uniform magnetic field, $B(r)$, where $r = \sqrt{x^2 + y^2}$ is the distance from the null point. The wave propagates in the $r$ direction. By symmetry, the wave perturbs the strength of the field, plasma density and pressure, and causes plasma flows in the $r$ direction. The wave is purely longitudinal, as the induced flows are parallel to the wave vector. All perturbed quantities depend upon time $t$ and $r$ only. For simplicity, let us restrict ourselves to isothermal perturbations, with the adiabatic index $\gamma = 1$. In the 2D equilibrium considered, the force caused by the gradient of the total pressure is counteracted by the magnetic tension force. In the adopted 1D model, the latter force does not affect the fast wave dynamics, as we neglect effects of the obliqueness.  

Linearising MHD equations near the equilibrium, we obtain the following wave equation,
\begin{equation}
\label{Eq:wave1}
\frac{\partial^2 V}{\partial t^2} - C_\mathrm{f}^2(r) \frac{\partial^2 V}{\partial r^2} = 
C_\mathrm{A}^2(r) \Big( 3 \frac{B'(r)}{B(r)} \frac{\partial V}{\partial r} + \frac{B''(r)}{B(r)} V \Big)  + {\cal N},
\end{equation}
where $V$ is the plasma velocity in the $r$-direction, and the prime denotes the derivative $d/d\,r$.
The Alfv\'en and fast speeds, $C_\mathrm{A}$ and $C_\mathrm{f}$, respectively, depend on $r$ due to the non-uniformity of the magnetic field. In addition, we include in Eq.~(\ref{Eq:wave1}) a nonlinear term ${\cal N}$ which will be omitted until its discussion in Section~\ref{sec:nl}.

If the field strength increases linearly with $r$, $B= B_{0} r/ L_0$, see Eq.~(\ref{Eq:beq}), the expressions for the characteristic speeds become
\begin{equation}
\label{Eq:splin}
\displaystyle C_\mathrm{A} = \frac{B_0 r}{L_0 \sqrt{4\pi \rho_0} }, \mbox{\ \ \ } C_\mathrm{f} = \sqrt{C_\mathrm{s}^2 + \frac{B_0^2 r^2}{L_0^2 {4\pi \rho_0} }},
\end{equation}
and Eq.~(\ref{Eq:wave1}) reduces to 
\begin{equation}
\label{Eq:wave2}
\displaystyle \frac{\partial^2 V}{\partial t^2} - \Big(C_\mathrm{s}^2 + \frac{B_0^2 }{L_0^2 {4\pi \rho_0} }r^2\Big) \frac{\partial^2 V}{\partial r^2} = 
\frac{3 B_0^2}{4\pi \rho_0 L_0^2}r \frac{\partial V}{\partial r}.
\end{equation}
Let $C_\mathrm{A0} = C_\mathrm{A}(r=L_0)$. 
In normalised independent variables, $\tau = t/\tau_0$ and $s = r/L_0$, where $\tau_0$ is the characteristic time scale, $\tau_0 = L_0/C_\mathrm{A0}$, Eq.~(\ref{Eq:wave2}) becomes
\begin{equation}
\label{Eq:wave3}
\displaystyle \frac{\partial^2 V}{\partial \tau^2} - \Big(\beta_0 + s^2\Big) \frac{\partial^2 V}{\partial s^2} = 3s \frac{\partial V}{\partial s},
\end{equation}
where $\beta_0 = C_\mathrm{s}^2/C_\mathrm{A0}^2$.  

\begin{figure}
    \centering
    \includegraphics[width=0.72\linewidth, trim={3 3 3 0}, clip]{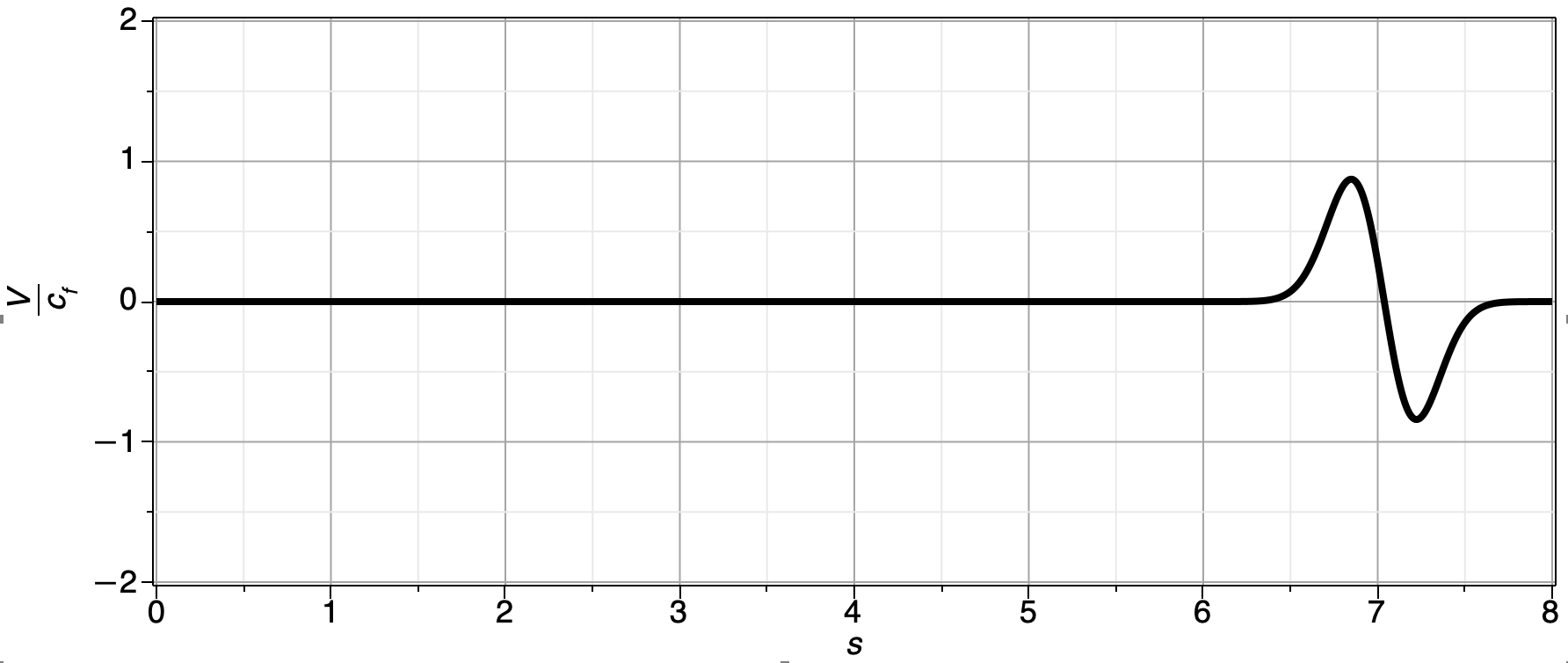}\\
    \includegraphics[width=0.72\linewidth, trim={3 3.3 3 0}, clip]{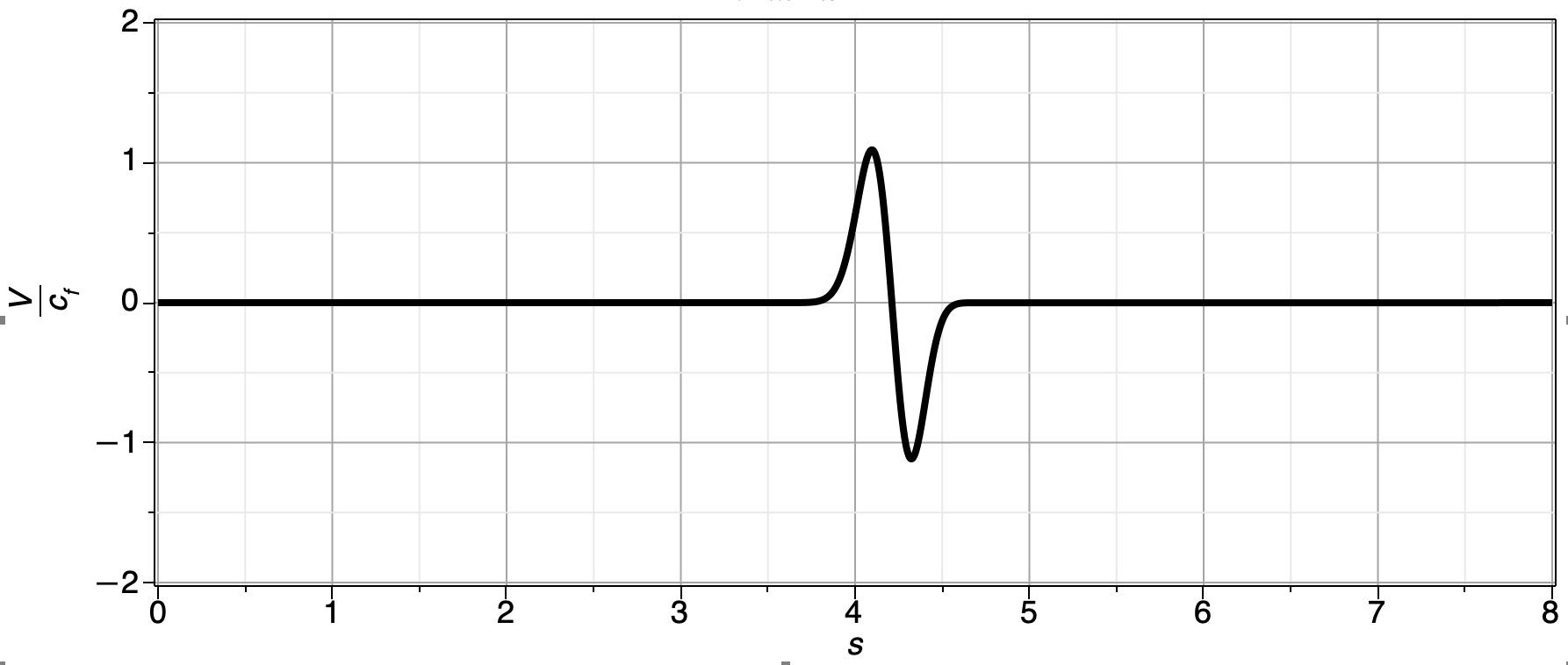}\\
    \includegraphics[width=0.72\linewidth, trim={0 3 3 11}, clip]{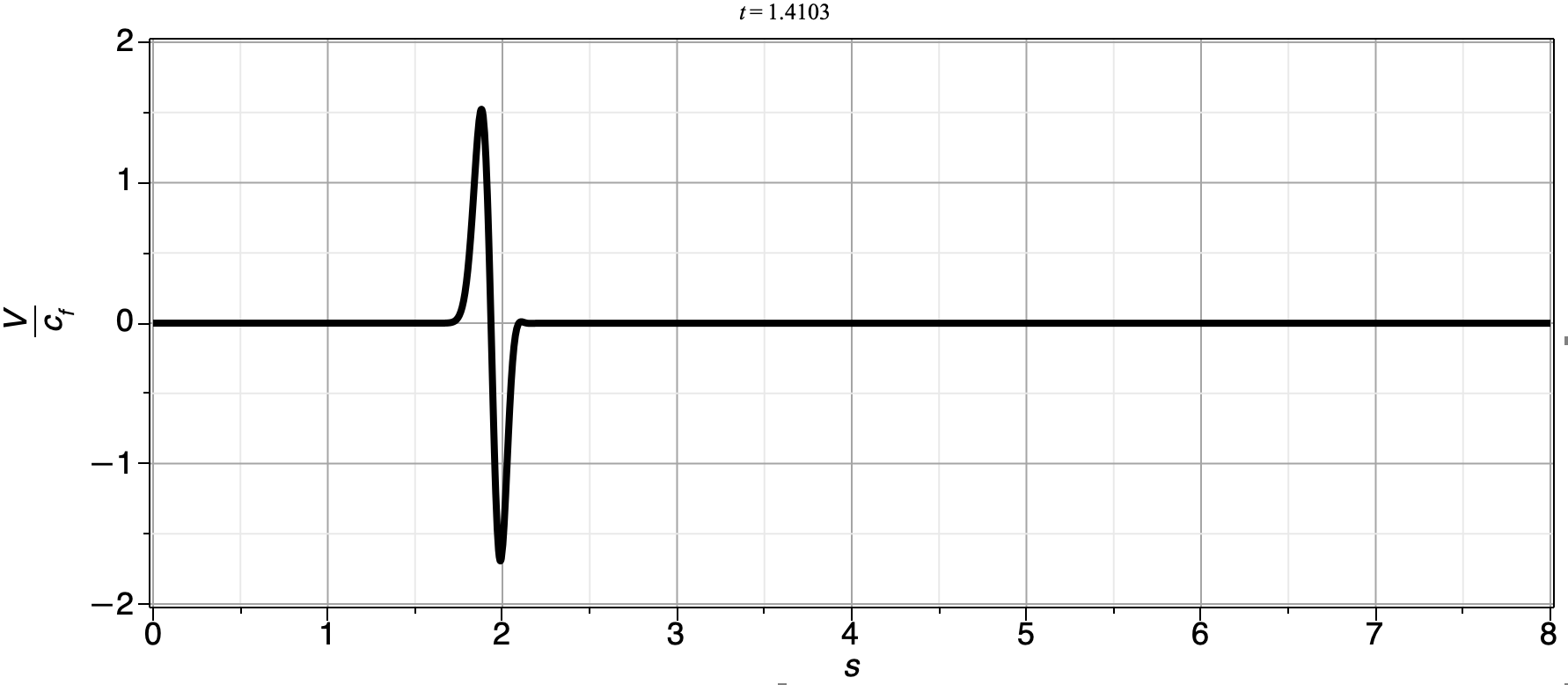}
     \caption{Evolution of a locally-plane linear fast wave approaching a null point along a magnetic bisector in the decreasing $s$ direction, i.e., from right to left. The figure shows the perpendicular component of the velocity, normalised to the local fast speed at the instants of time $\tau=0.13$ (top), $\tau=0.64$ (middle), and $\tau = 1.41$ (bottom). The parameter $\beta_0 = 0.2$.}
    \label{fig:an1}
\end{figure}

Eq.~(\ref{Eq:wave3}) could be solved analytically, but it is easier to look at its numerical solutions which can be readily obtained with the function \textit{pdsolve} of the Maple computing environment\footnote{In this study, we used the Maple 2024.2 version.}.
Figure~\ref{fig:an1} shows the evolution of a wave pulse, initially shaped as the derivative of a Gaussian, propagating toward the null point. As expected, the wavelength decreases, while its amplitude grows. The density perturbation $\rho_1$ behaves in a similar manner, according to the continuity equation, $V/C_\mathrm{f} \approx \rho_1/\rho_0$. 

The increase in the amplitude and decrease in the wavelength lead to the decrease in the shock formation distance, see, e.g., \cite{2011A&A...531A..63G}, which is addressed in Section~3\ref{sec:nl}. 

\begin{figure}
    \centering
    \includegraphics[width=0.6\linewidth]{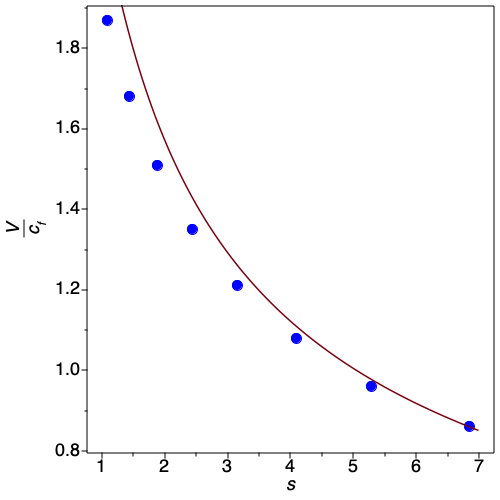}
     \caption{Comparison of the amplitude of a locally-plane linear fast wave approaching a null point along a magnetic bisector, obtained in the WKB approximation (curve) and numerically with the full equation (circles). The parameter $\beta_0 = 0.2$. }
    \label{fig:an2}
\end{figure}

\subsection{Short-wavelength limit}
\label{sec:wkb}

If the effective wave length is much shorter than the size of the non-uniformity $L$, we can employ the  Wentzel--Kramers--Brillouin (WKB) approximation. In this case, one can neglect the right hand side terms in Eq.~(\ref{Eq:wave3}), reducing it to
\begin{equation}
\label{Eq:wave4}
\displaystyle \frac{\partial^2 V}{\partial \tau^2} - \Big(\beta_0 + s^2\Big) \frac{\partial^2 V}{\partial s^2} = 0.
\end{equation}

A solution to Eq.~(\ref{Eq:wave4}) that describes a wave propagating in the negative \added{$s$} direction is
\begin{equation}
\label{Eq:wkb}
   V(\tau, s) \approx \Big(\beta_0 + s^2\Big)^{1/4} F \Big(\tau +\mathrm{arcsinh}(s/\sqrt{\beta_0})\Big),
\end{equation} 
where the function $F$ describes the shape of the wave (see, e.g. \cite{1974lnw..book.....W}). In the zero-$\beta$ limit, it reduces to
\begin{equation}
\label{Eq.wkb0}
   V(\tau, s) \approx s^{1/2} F \Big(\tau + \mathrm{ln}(s)\Big).
\end{equation}
The same result has been obtained in \cite{2006A&A...459..641M}.

The wave propagating in the negative $s$ direction toward $s=0$ gradually slows down, and its amplitude decreases. However, its relative amplitude increases,
\begin{equation}
\label{Eq:wkbr}
\frac{V}{C_\mathrm{f}(s)} \propto \Big(\beta_0 + s^2\Big)^{-1/4}.
\end{equation}
For $\beta_0 = 0$, this expression reduces to
\begin{equation}
\label{Eq:wkbr0}
\frac{V}{C_\mathrm{A}(s)} \propto s^{-1/2}.
\end{equation}
This behaviour is qualitatively consistent with the numerical solution of the full version of Eq.~(\ref{Eq:wave3}), see Figure~\ref{fig:an2}.

\subsection{Weakly nonlinear short-wavelength limit}
\label{sec:nl}

Assuming the wave amplitude to be finite but small, we can restrict our attention to quadratically nonlinear terms only.
In the WKB approximation, we can neglect the spatial derivatives of equilibrium quantities in the nonlinear terms. In addition, we neglect finite-$\beta$ terms,  obtaining  
\begin{equation}
\label{Eq:nl1}
{\cal N} = -\frac{1}{\rho_0}\frac{\partial}{\partial \tau}\Big(\rho_1 \frac{\partial V}{\partial \tau}\Big) - \frac{1}{C_\mathrm{A0}}\frac{\partial}{\partial \tau}\Big(V\frac{\partial V}{\partial s}\Big) -\frac{1}{4\pi \rho_0 C_\mathrm{A0}}\frac{\partial}{\partial \tau}\Big(B_1\frac{\partial B_1}{\partial s}\Big) + \frac{B_0}{4\pi \rho_0 C_\mathrm{A0}^2}\frac{\partial^2 }{\partial s^2}\Big(B_1 V\Big).
\end{equation}
Using the linear part of the continuity equation and induction equation, we express the liner perturbation of the density and magnetic field via the linear velocity,
\begin{equation}
 \rho_1 \approx \rho_0\, V/C_\mathrm{f}, \mbox{\ \ \ } B_1 \approx B_0\, V/C_\mathrm{f}, \mbox{\ \ \ }  C_\mathrm{f} = C_\mathrm{A0}s,
\end{equation}
which allows us to exclude $\rho_1$ and $B_1$ from the nonlinear terms (\ref{Eq:nl1}). Thus, we modify Eq.~(\ref{Eq:wave4}) as
% \begin{equation}
% \label{Eq:wave5}
% \displaystyle \frac{\partial^2 V}{\partial \tau^2} - s^2 \frac{\partial^2 V}{\partial s^2} = -\frac{1}{C_\mathrm{A0}s}\frac{\partial}{\partial \tau} \Big(V \frac{\partial V}{\partial \tau}\Big) - \frac{1}{C_\mathrm{A0}}\frac{\partial}{\partial \tau}\Big(V\frac{\partial V}{\partial s} \Big)- \frac{1}{C_\mathrm{A0} s^2}\frac{\partial}{\partial \tau}\Big(V\frac{\partial V}{\partial s} -\frac{V^2}{s^2} \Big)
% \end{equation}
% $$
% \displaystyle
% \mbox{\ \ \ \ \ \ \ \ \ \ \ \ \ \ \ \ \ \ \ \ \ \ \ \ \ \ \ }
%  +\frac{1}{C_\mathrm{A0}}\frac{\partial^2}{\partial s^2}\Big( \frac{V^2}{s} \Big).
% $$
\begin{align}
\label{Eq:wave5}
\displaystyle \frac{\partial^2 V}{\partial \tau^2} - s^2 \frac{\partial^2 V}{\partial s^2} &= -\frac{1}{C_\mathrm{A0}s}\frac{\partial}{\partial \tau} \Big(V \frac{\partial V}{\partial \tau}\Big) - \frac{1}{C_\mathrm{A0}}\frac{\partial}{\partial \tau}\Big(V\frac{\partial V}{\partial s} \Big)- \frac{1}{C_\mathrm{A0} s^2}\frac{\partial}{\partial \tau}\Big(V\frac{\partial V}{\partial s} -\frac{V^2}{s^2} \Big) \nonumber \\
    &\quad +\frac{1}{C_\mathrm{A0}}\frac{\partial^2}{\partial s^2}\Big( \frac{V^2}{s} \Big). 
\end{align}
%Under the weakly non-uniform background assumption, the characteristic scale of the background variation is much larger than the wavelength, which allows us to 

The assumptions of the weak non-linearity and non-uniformity allow us to apply the method of the slowly varying amplitude (see, e.g., \cite{1974lnw..book.....W, 2011A&A...531A..63G}). Let us introduce new independent variables
\begin{equation}
    \displaystyle \xi = \tau +\int \frac{ds}{s}  = 
    \tau + \mathrm{ln}(s),\mbox{\ \ \ \ } R=\epsilon s
\end{equation}
where $\xi$ is the spatial coordinate associated with the wave propagating in the negative $s$ direction, and $\epsilon$ denotes a small parameter characterising the smallness of the nonlinear term, i.e., of the wave amplitude. Then, Eq.~(\ref{Eq:wave5}) is rewritten as
\begin{equation}
    \displaystyle   \frac{\partial V}{\partial \xi} -2 R \frac{\partial^2 V}{\partial \xi\partial R} - R^2 \frac{\partial^2 V}{\partial R^2}   = -\frac{\epsilon}{C_\mathrm{A_0}}\frac{\partial}{\partial \xi}\left(\frac{2}{R}V\frac{\partial V}{\partial \xi} +  V \frac{\partial V}{ \partial R} \right) + \mathcal{O}(\epsilon^2).
\end{equation}
Neglecting higher–order terms in $\epsilon$ and assuming $V(\xi, R)=A(R)U(\xi, R)$ allows us to separate the fast-varying phase and the slowly-varying envelope. Taking into account that the derivatives with respect to $R$ are much smaller than with respect to $\xi$,  we obtain the equation 
\begin{equation}
\label{Eq:wave6}
    \displaystyle  -2R\frac{\partial^2 U}{\partial \xi \partial R} = -\frac{\partial}{\partial \xi}\Big(\frac{2\epsilon A}{C_\mathrm{A0}R}U\frac{\partial U}{\partial \xi}\Big).   
%    + \frac{\epsilon A}{C_\mathrm{A0}}U\frac{\partial U}{\partial R} + \frac{\epsilon}{C_\mathrm{A0}}\frac{\partial A}{\partial R}U^2 \Big).
\end{equation}
Eq.~(\ref{Eq:wave6}) can be readily integrated with respect to $\xi$, giving us 
\begin{equation}
\label{Eq:wave7}
    \frac{\partial U}{ \partial R} = \frac{\epsilon A(R)}{C_\mathrm{A0}R^2}U\frac{\partial U}{\partial \xi}.
\end{equation}
By substituting $A(s) \propto s^{1/2}$, as established in Eq.~(\ref{Eq:wkbr0}), and by absorbing $\sqrt{\epsilon}$ into $\tilde{U}$, Eq.~(\ref{Eq:wave7}) can be rewritten as
\begin{equation}
\label{Eq:wave8}
    \frac{\partial \tilde{U}}{ \partial R} - \frac{1}{C_\mathrm{A0} R^{3/2}}\tilde{U}\frac{\partial \tilde{U}}{\partial \xi} =0.
%    \ \textit{(or,} \frac{\partial U}{ \partial R} - \frac{A(R_0)\sqrt{\epsilon}}{C_\mathrm{A0} R^{3/2}}U\frac{\partial U}{\partial \xi} =0, \epsilon = \lambda/L_0 \textit{)}
\end{equation}
This equation is an inviscid Burgers equation, also known as a simple wave equation, with a non-uniform coefficient. It represents the planar analogue of Eq.~(25) derived in \cite{2011A&A...531A..63G}. 

Eq.~(\ref{Eq:wave8}) describes the wave steepening (e.g., \cite{1974lnw..book.....W}), and allows us to link the shock formation distance with parameters of the wave. Consider the evolution of a single harmonic which at a distance $R=R_0$ is given by the expression
\begin{equation}
\label{Eq:harm}
\tilde{U}(\xi,R_0) = A_0 \cos(k_0 \xi),
\end{equation}
where $A_0$ is the initial amplitude and $k_0$ the wave number. 

As Eq.~(\ref{Eq:wave8}) is of the first order, we can determine its characteristics,
\begin{equation}
\label{Eq:chars1}
\displaystyle \frac{d\, \tilde{U}}{d\,R}=0, \mbox{\ \ \ \ } \frac{d\,\xi}{d\,R}=  - \frac{A_0\cos(k_0\xi_0)}{C_\mathrm{A0} R^{3/2}},
\end{equation}
where $\xi_0$ is the initial location of a certain phase of the wave.
%we used that $U$ is constant along the characteristic.
By separating the variables and integrating, we obtain
\begin{equation}
\label{Eq:chars2}
\xi(\xi_0,R) = \xi_0 + \frac{2A_0\cos(k_0\xi_0)}{C_\mathrm{A0}}\big( R^{-1/2} - R_0^{-1/2}\big).
\end{equation}
A shock forms when characteristics intersect, i.e., 
\begin{equation}
\label{Eq:chars3}
\frac{\partial \xi}{\partial \xi_0}
= 1 - \frac{2A_0 k_0 \sin(k_0\xi_0) }{C_\mathrm{A0}}\big(R^{-1/2}-R_0^{-1/2}\big) = 0.
\end{equation}
The earliest shock occurs at $\sin(k_0\xi_0) = 1$, hence the shock formation distance is
\begin{equation}
\label{Eq:chars4}
R_\mathrm{sf} = \left(R_0^{-1/2} + \frac{C_\mathrm{A0}}{2 A_0 k_0}\right)^{-2}.
\end{equation}
As in the uniform medium, the shock forms earlier, i.e., $R_\mathrm{sf}$ is larger, for larger initial amplitudes $A_0$ and higher spatial harmonics $k_0$. For non-harmonic waves, the dependence of the shock formation distance on the wave number means that the shock forms at steeper parts of the wave.
Eq.~(\ref{Eq:chars4}) indicates that incoming waves with larger amplitudes are subject to nonlinear dissipation at larger distances from the null point, i.e., larger $R_\mathrm{sf}$. 

\section{2D MHD numerical solutions}
\label{sec:num}

We utilised FLASH code \cite{2010ascl.soft10082F} to numerically solve the set of MHD equations on an adaptive mesh refinement grid with the unsplit staggered mesh solver. To accurately capture fine-scale wave dynamics with minimal numerical dissipation, the Harten--Lax--van Leer\added{--}discontinuities \added{(HLLD)} Riemann solver is used alongside a second-order monotonic upstream-centred scheme for conservation laws for data reconstruction. \added{The highest resolution in simulation corresponds to cells size of approximately $46.8\ \mathrm{km}$ within a physical domain of $240\ \mathrm{Mm} \times240\ \mathrm{Mm} $.}%To mitigate numerical instabilities at grid resolution interfaces, a controllable artificial viscosity is introduced.

An MHD wave perturbation is initiated by a circular velocity pulse,
\begin{equation}
\begin{aligned}
    \pmb{v}_x &= -A_0 \cos(\theta)\cos\left( \pi \frac{\sqrt{(x-x_0)^2 + (y-y_0)^2}}{2w_\mathrm{0}} \right), \\
    \pmb{v}_y &= -A_0 \sin(\theta)\cos\left( \pi \frac{\sqrt{(x-x_0)^2 + (y-y_0)^2}}{2w_\mathrm{0}} \right),
\end{aligned}
\end{equation}
where $A_0$, $\theta$ and $w_\mathrm{0}$ represent the initial velocity amplitude, the azimuth angle $\theta = \arctan(y_0/x_0)$ of the pulse centre $(x_0, y_0)$, and the radius of the pulse region, respectively. The centre of the initial pulse is located at the distance $r_0 = \sqrt{x_0^2+y_0^2}$ from the origin. 

The evolution of the magnetoacoustic pulse in the vicinity of the magnetic null point is modelled for the following combination of parameters: $\rho_0 = 8.36\times 10^{-16}$~g/cm$^{3}$, $T_0 = 10^6$~K, $B_0 = 3$~G, $L_0 = 20$~Mm, and $(x_0,\ y_0 )= (20,\ 20)$~Mm which corresponds to the distance $r_0 \approx 28.3$~Mm from the null point, see Figure~\ref{fig:xp}. The sound speed in the computational domain is $C_\mathrm{s}=166$~km\,s$^{-1}$, and the Alfv\'en speed is $C_\mathrm{A}(r_0)=415$~km\,s$^{-1}$ at the location of the initial pulse. The initial location is outside the circle $C_\mathrm{A} = C_\mathrm{s}$. The amplitude $A_\mathrm{i}$ and width $w_\mathrm{0}$ of the initial pulse vary. 

\begin{figure}
    \centering
    \includegraphics[width=0.45\linewidth, trim={30 30 30 10}, clip]{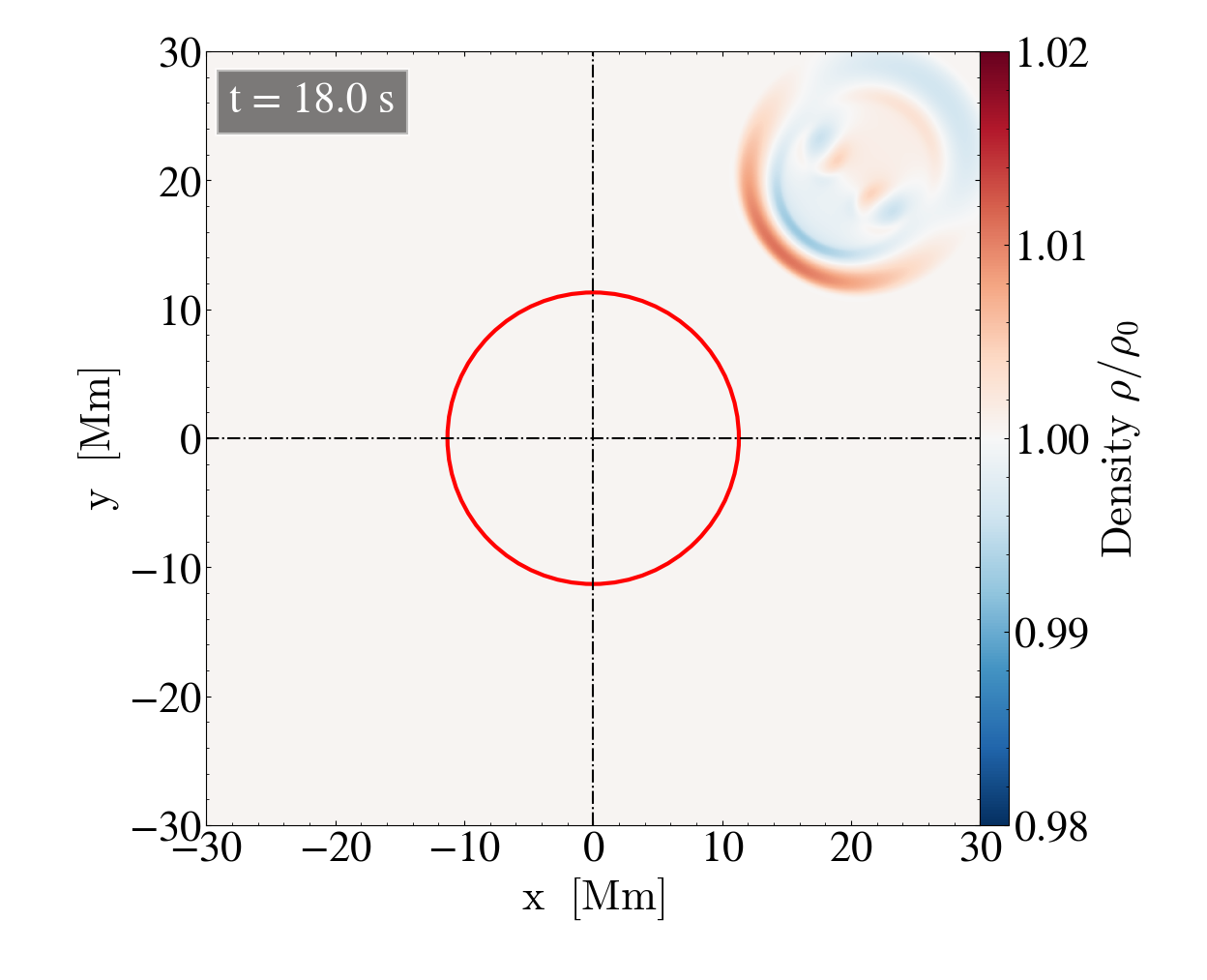}
    \includegraphics[width=0.45\linewidth, trim={30 30 30 10}, clip]{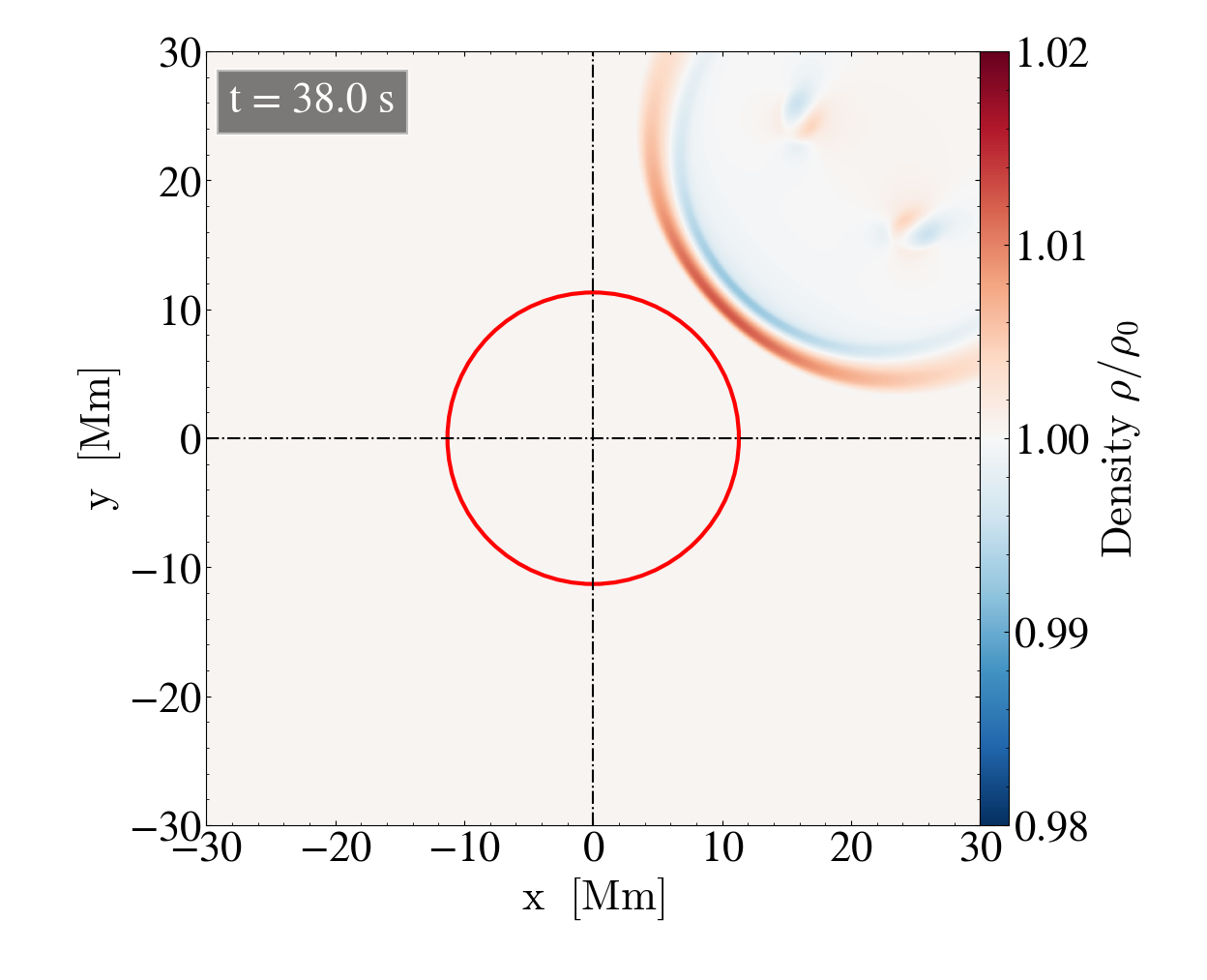}
    \includegraphics[width=0.45\linewidth, trim={30 30 30 10}, clip]{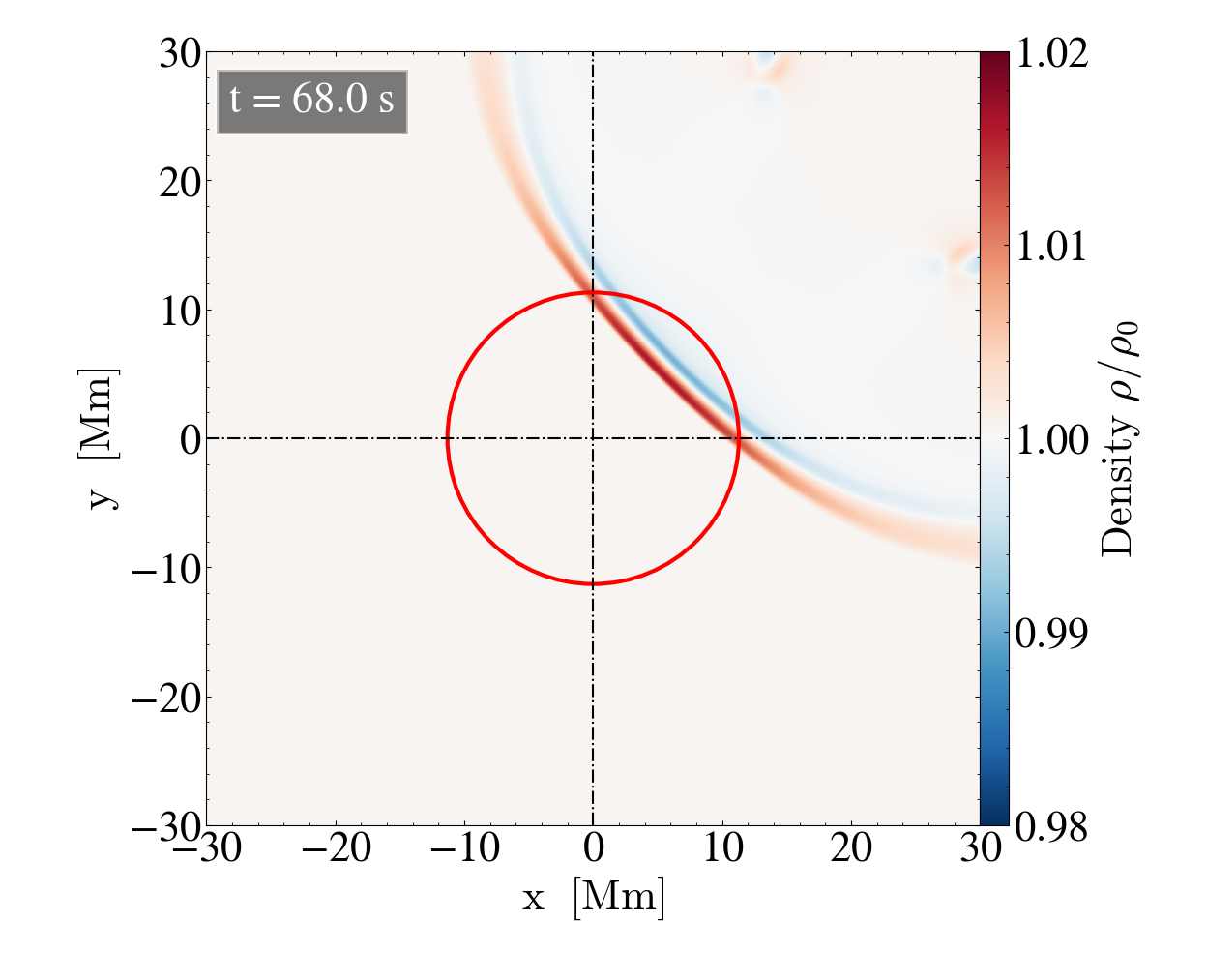}
    \includegraphics[width=0.45\linewidth, trim={30 30 30 10}, clip]{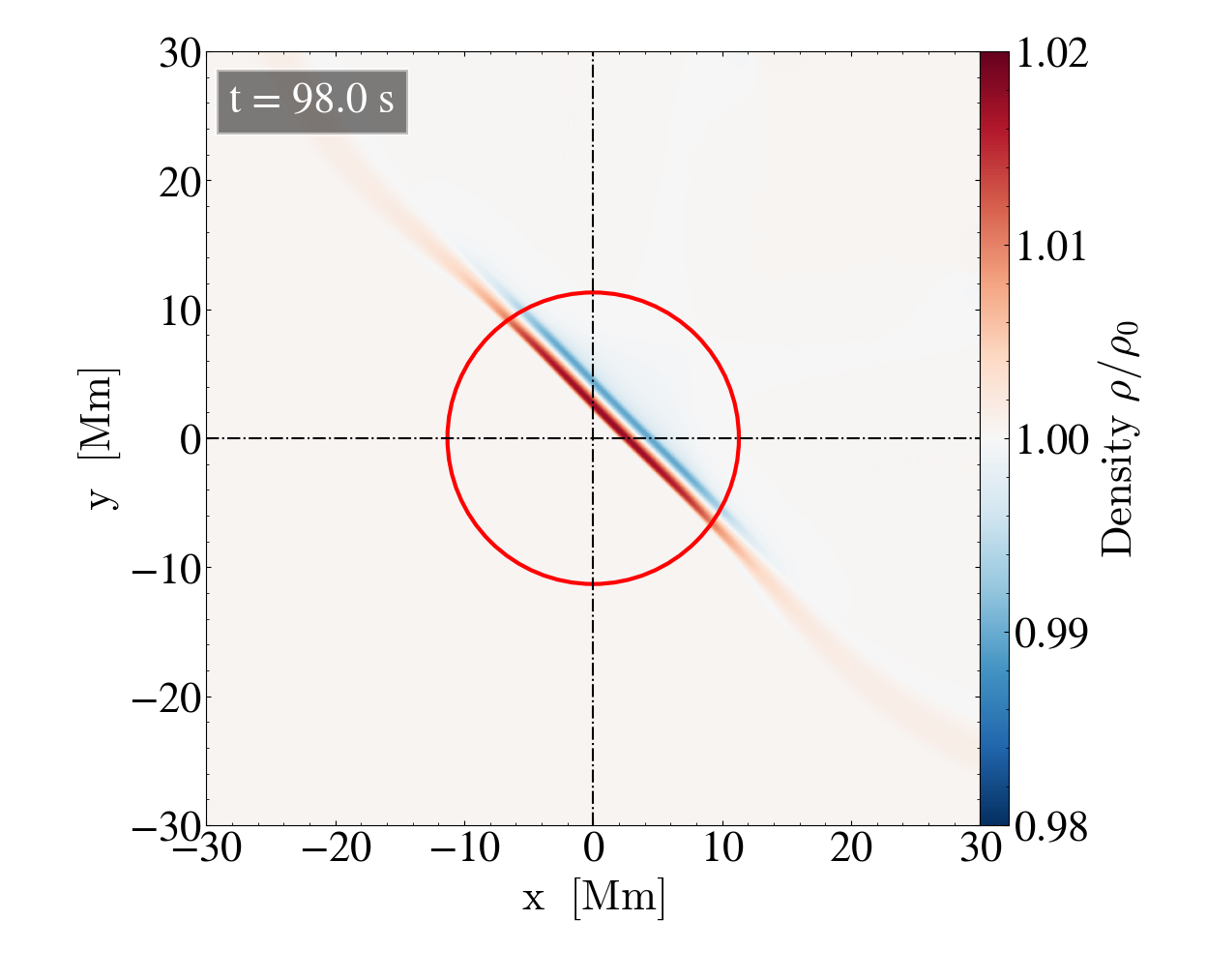}
    \caption{Perturbations of the density in a fast wave pulse excited by a remote localised source, approaching a 2D magnetic null point. The red circle indicates the $C_\mathrm{A} = C_\mathrm{s}$ distance. The snapshots are taken at the instants of time, indicated in the grey inlets. }
    \label{fig:2d}
\end{figure}

\subsection{Evolution of the pulse}

Figure~\ref{fig:2d} demonstrates the typical evolution of an incoming fast wave pulse in the vicinity of a null point. The initial perturbation splits into slow waves that propagate almost along the local magnetic field, and a fast wave which propagates obliquely and perpendicular to the field. Both excited waves are magnetoacoustic and hence perturb the density of the plasma. Gradually, the initially circular, in the 2D geometry, fast wave front experiences deformation. The fast wave front segment which propagates across the equilibrium field, becomes more planar. In this study, we are interested in the evolution of this part of the fast wave pulse.

\begin{figure}
    \centering
    \includegraphics[width=0.462\linewidth, trim = {0 0 0 0}, clip]{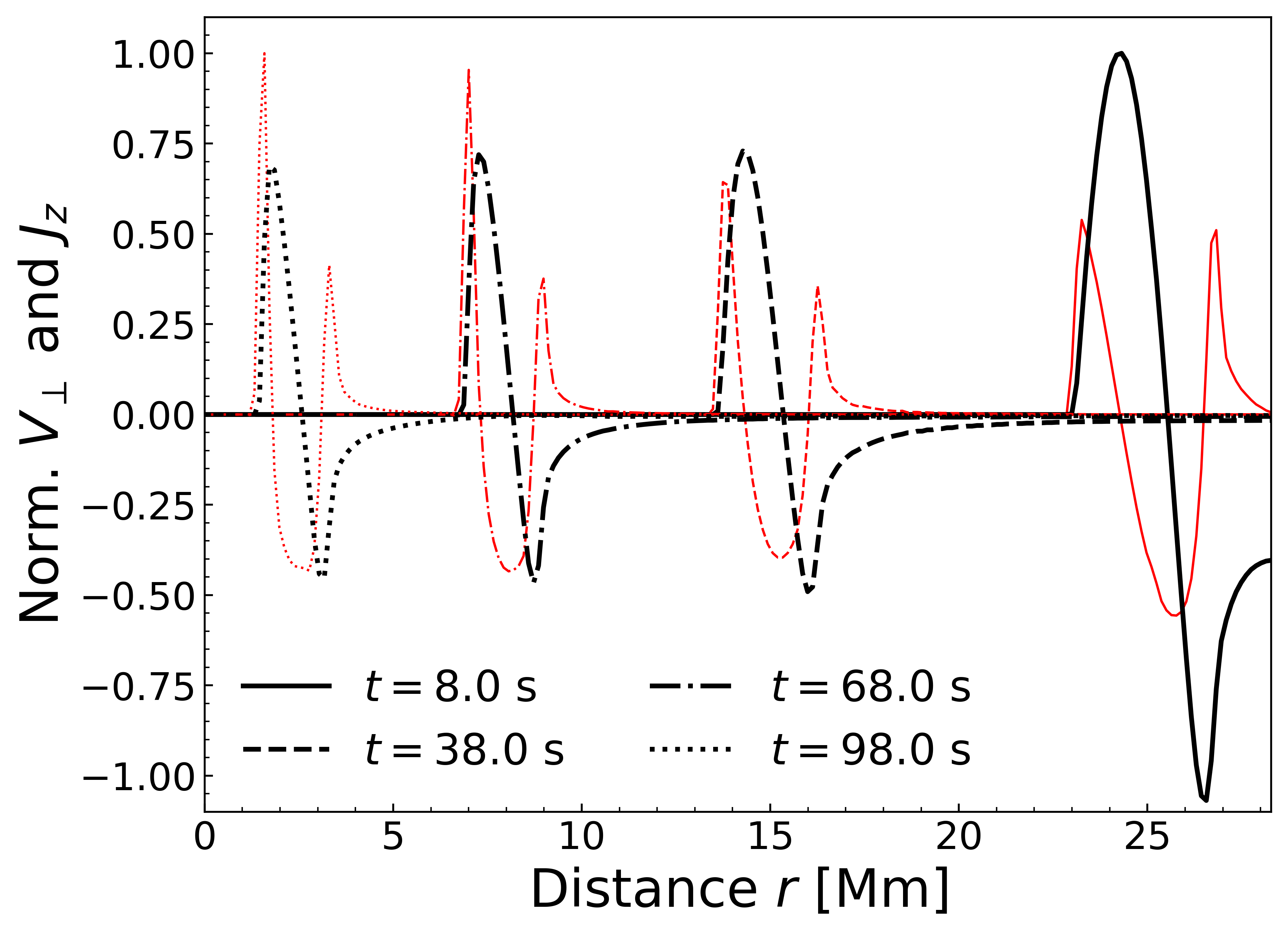}
    \includegraphics[width=0.52\linewidth, trim = {0 0 0 0}, clip]{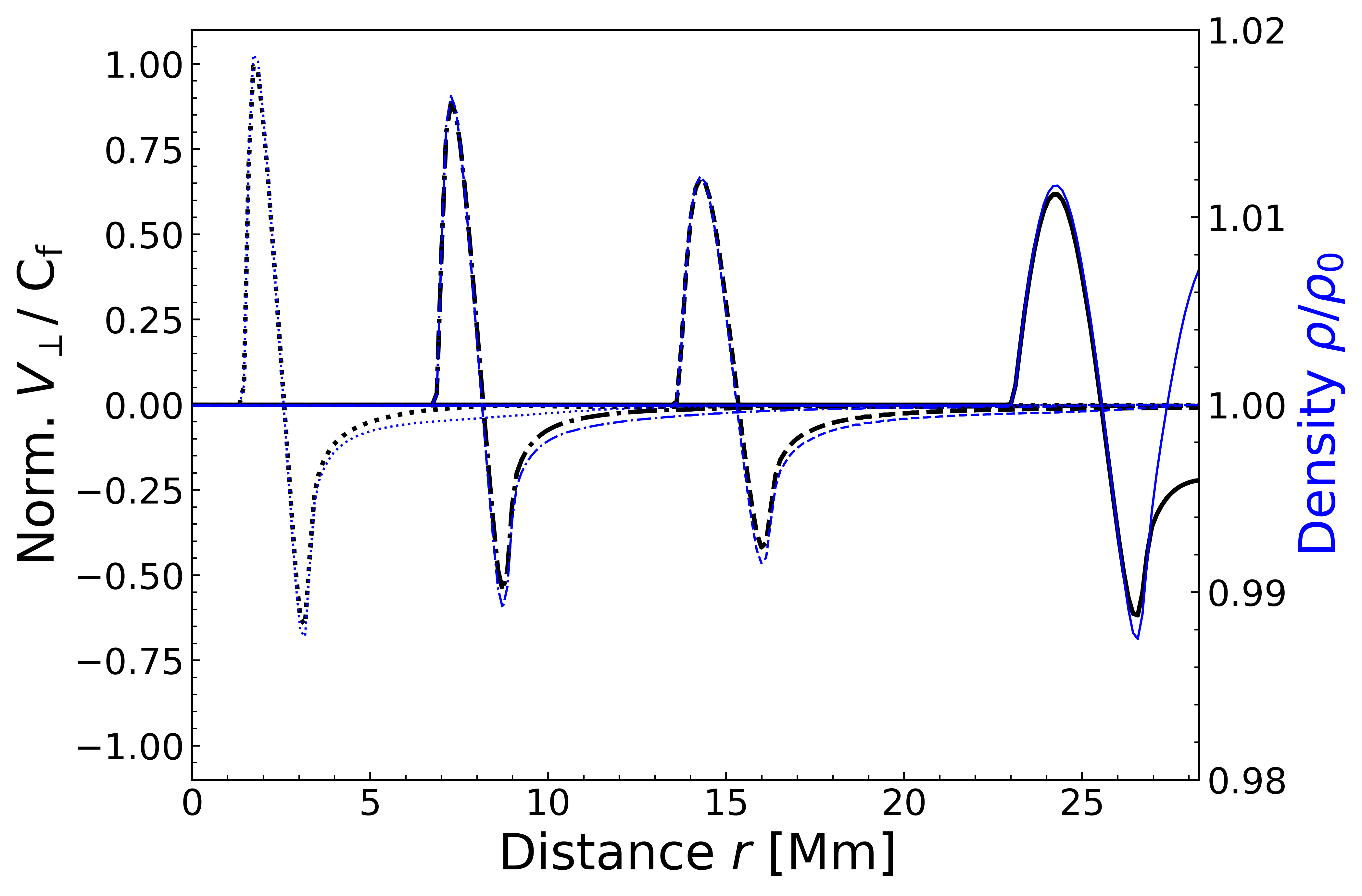}
    \includegraphics[width=0.462\linewidth, trim = {0 0 0 0}, clip]{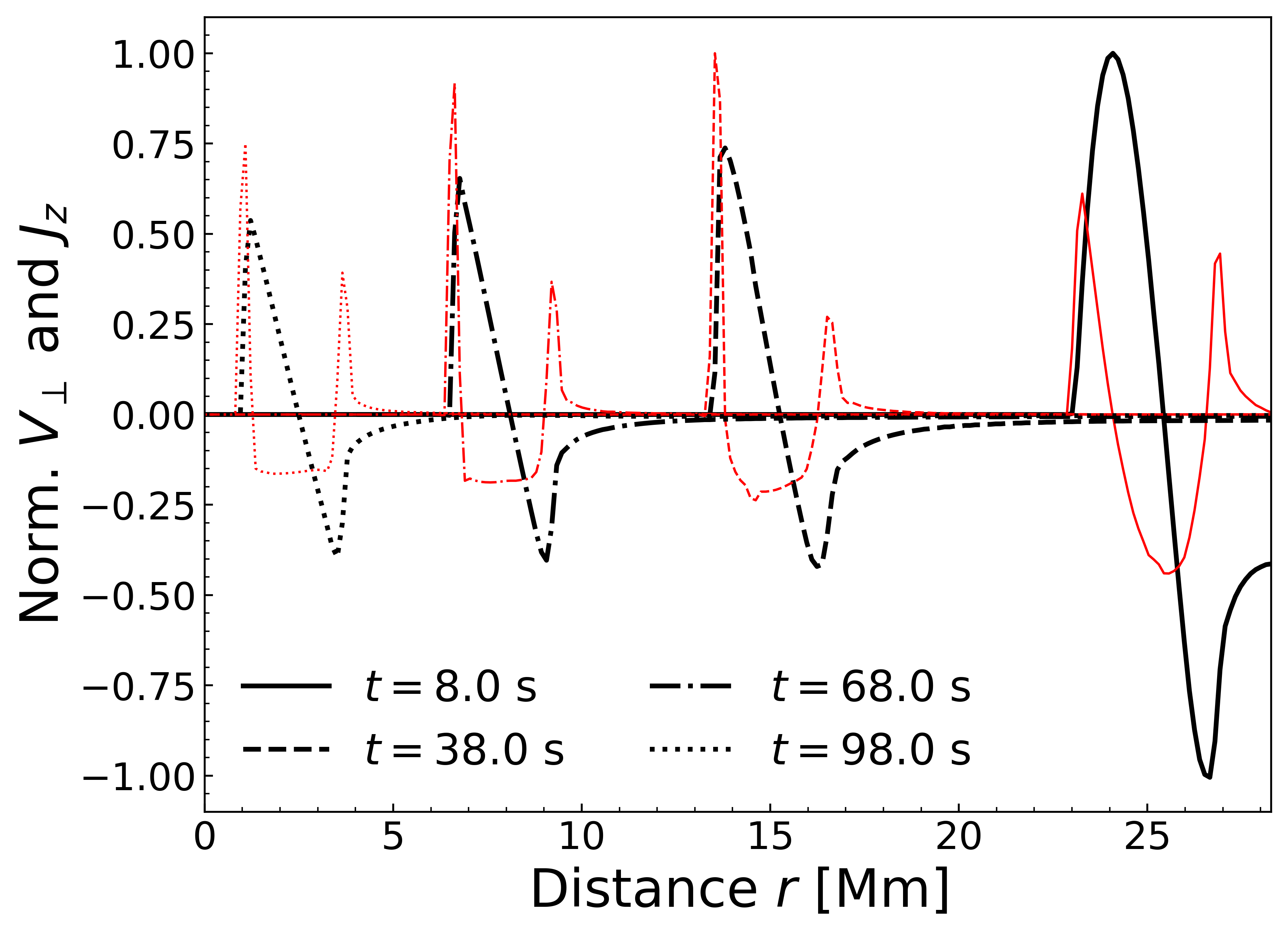}
    \includegraphics[width=0.52\linewidth, trim = {0 0 0 0}, clip]{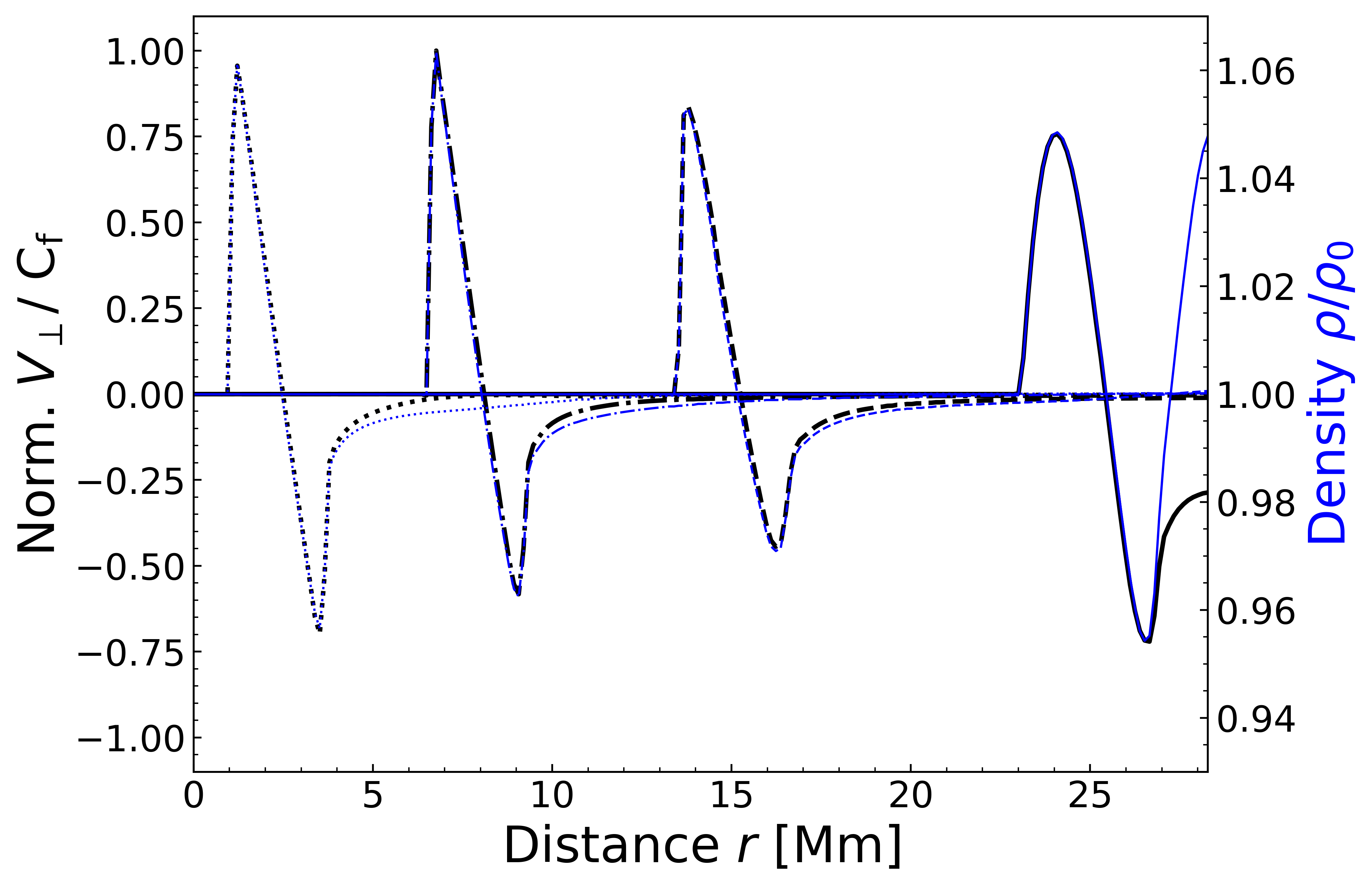}
    \includegraphics[width=0.462\linewidth, trim = {0 0 0 0}, clip]{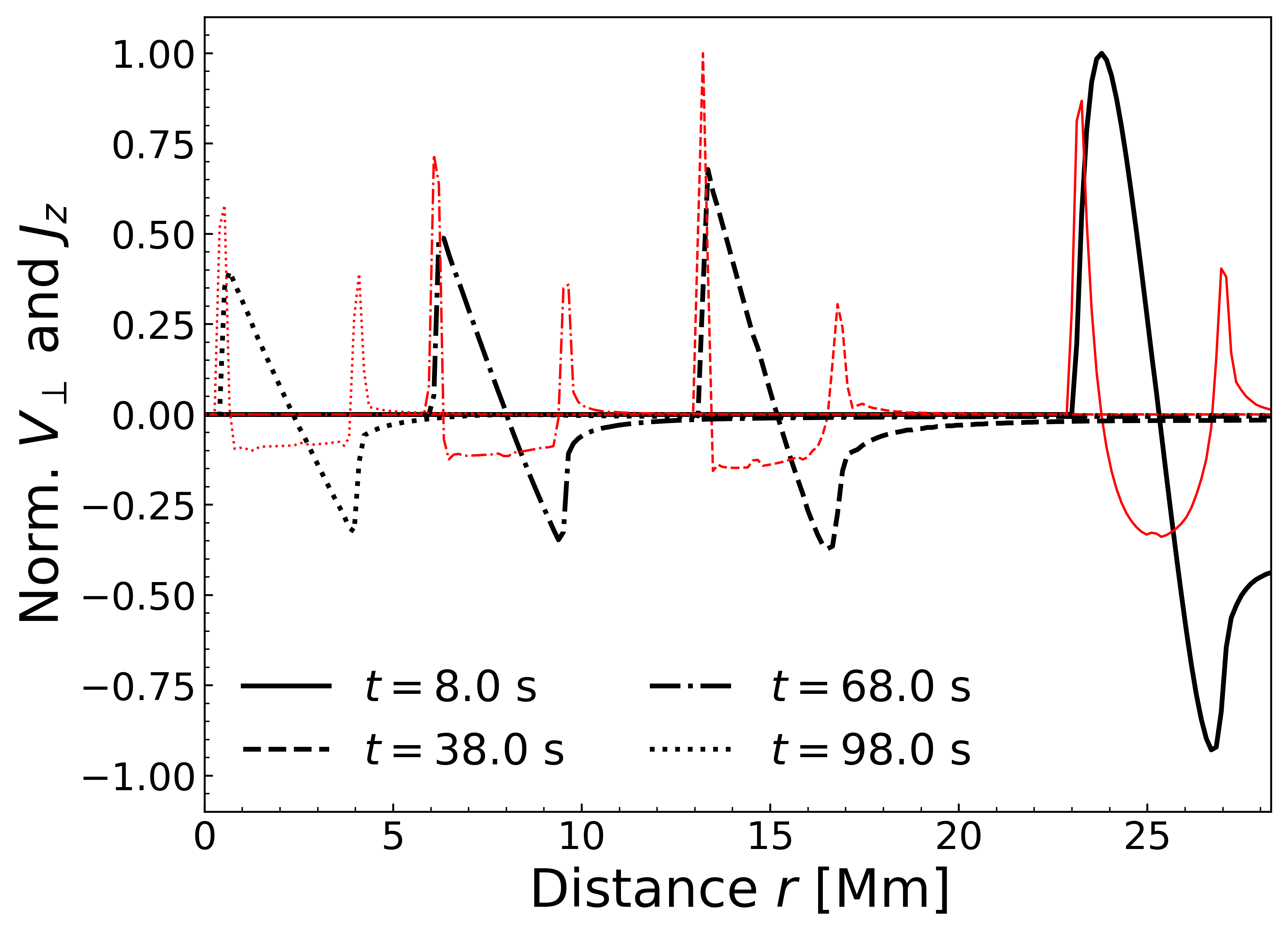}
    \includegraphics[width=0.52\linewidth, trim = {0 0 0 0}, clip]{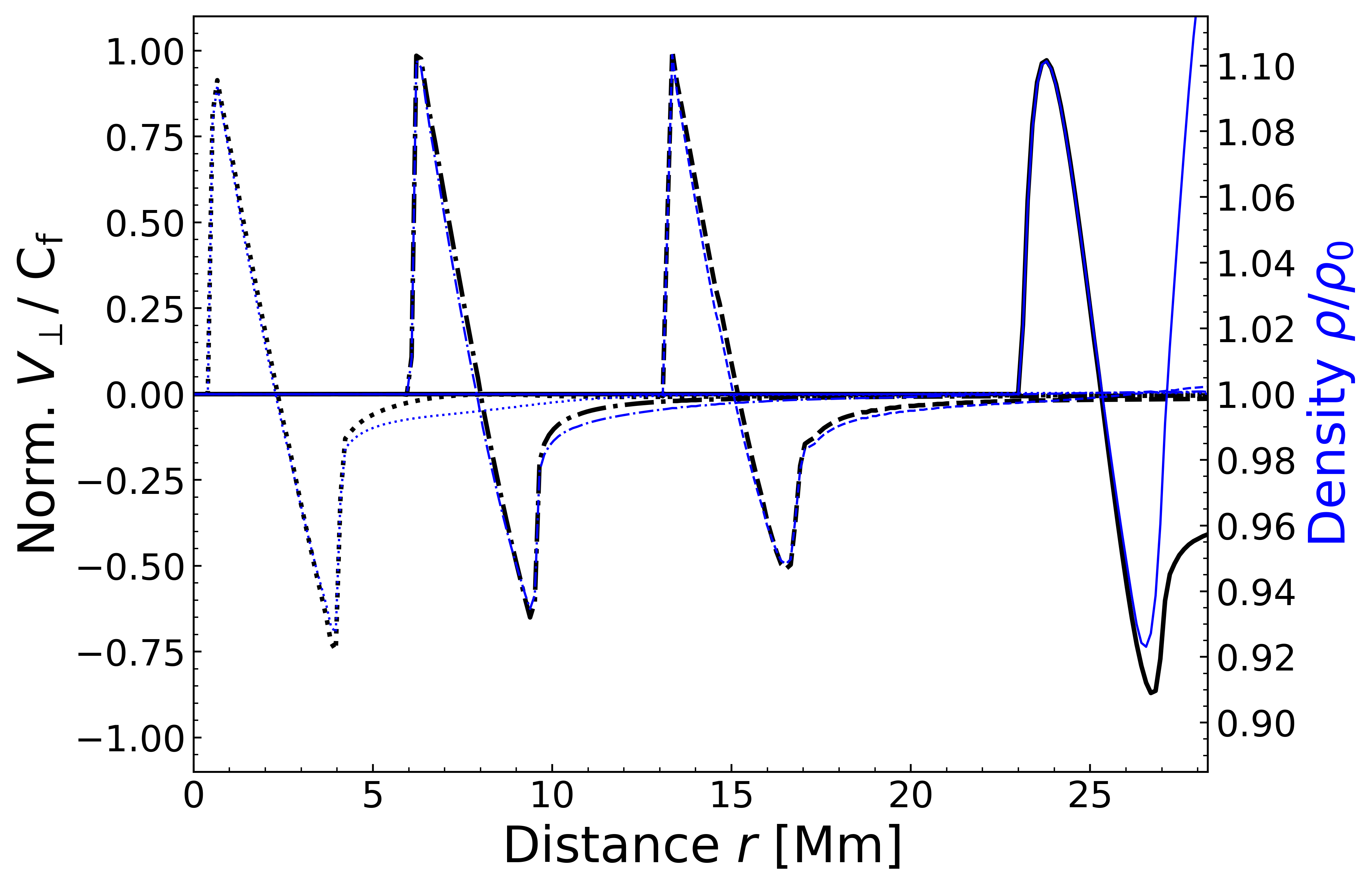}
    \caption{Evolution of the shape of the fast wave pulse with the distance from the null point. The left and right columns show, respectively, the perpendicular velocity and that velocity normalised to the local fast speed. The initial pulse has the width $w_0 = 2$~Mm, and the amplitude 20~\kms (upper row), 80~\kms (middle row) and 160~\kms (bottom row). The red \added{and blue} curves demonstrate the perturbation of the current density $J_z$ \added{($\mathrm{statA/cm^{2}}$) and density $\rho / \rho_0$} in the pulse\added{, respectively}. All curves are normalised to the maximum value. }
    \label{fig:vperp}
\end{figure}

Figure~\ref{fig:vperp} demonstrates the evolution of the shape of the fast pulse propagating along the magnetic bisector, i.e., across the unperturbed magnetic field. As it approaches the null point, the pulse becomes shorter. Its amplitude decreases, while the ratio of the amplitude to the local fast speed increases. This behaviour is in agreement with the analytical estimations made in a 1D approximation in Section~3\ref{sec:wkb}. 
Furthermore, the pulse is subject to nonlinear steepening, which occurs earlier, i.e., at a larger distance from the null point, for the pulses with larger amplitudes. Near the null point, the relative amplitude (right column) of the lower-amplitude pulse increases more rapidly than that of the higher-amplitude pulse. This discrepancy is attributed to dissipation in the vicinity of the shock, i.e. nonlinear damping that occurs after the shock formation, which is more effective for the higher amplitude. 

In our model, the governing equations do not include explicit dissipative terms, so the dissipation is numerical. This does not affect our results, since nonlinear damping is independent of the specific value of the dissipation coefficient, provided it is small (see, e.g., \cite{1974lnw..book.....W}). \added{Resolution tests showed that the shock steepening distance is independent of the chosen grid refinement level.} In the pulse of the considered shape, we see the formation of two shocks, on the leading and trailing slopes. In the vicinity of these shocks, spikes of the electric current density are observed. For waves of a lower initial amplitude, the current density spikes appear closer to the null point, whereas higher amplitude waves undergo more effective nonlinear steepening and are therefore \lq\lq overturned\rq\rq\ at a larger distance from the null point. 

The perpendicular fast magnetoacoustic wave is compressive: it perturbs the plasma density and the magnitude of the magnetic field, while leaving the field’s direction essentially unchanged. The perturbation of the field strength, $|\pmb{B}|$ is accompanied by a perturbation of the electric current density, $\pmb{j} = \frac{c}{4\pi}\nabla \times \pmb{B}$, which in our 2D model points out of the model plane. 
Steeper spatial gradients in the value of the magnetic field produce sharper spikes of the electric current density, so the strongest current concentrations occur at the shock fronts, see the left column of Figure~\ref{fig:vperp}. The true peak current densities are likely to be higher than shown, because the current spikes are not fully resolved and are artificially smoothed by numerical diffusion.

\subsection{Wave steepening distance}

\begin{figure}
    \centering
    \includegraphics[width=0.49\linewidth]{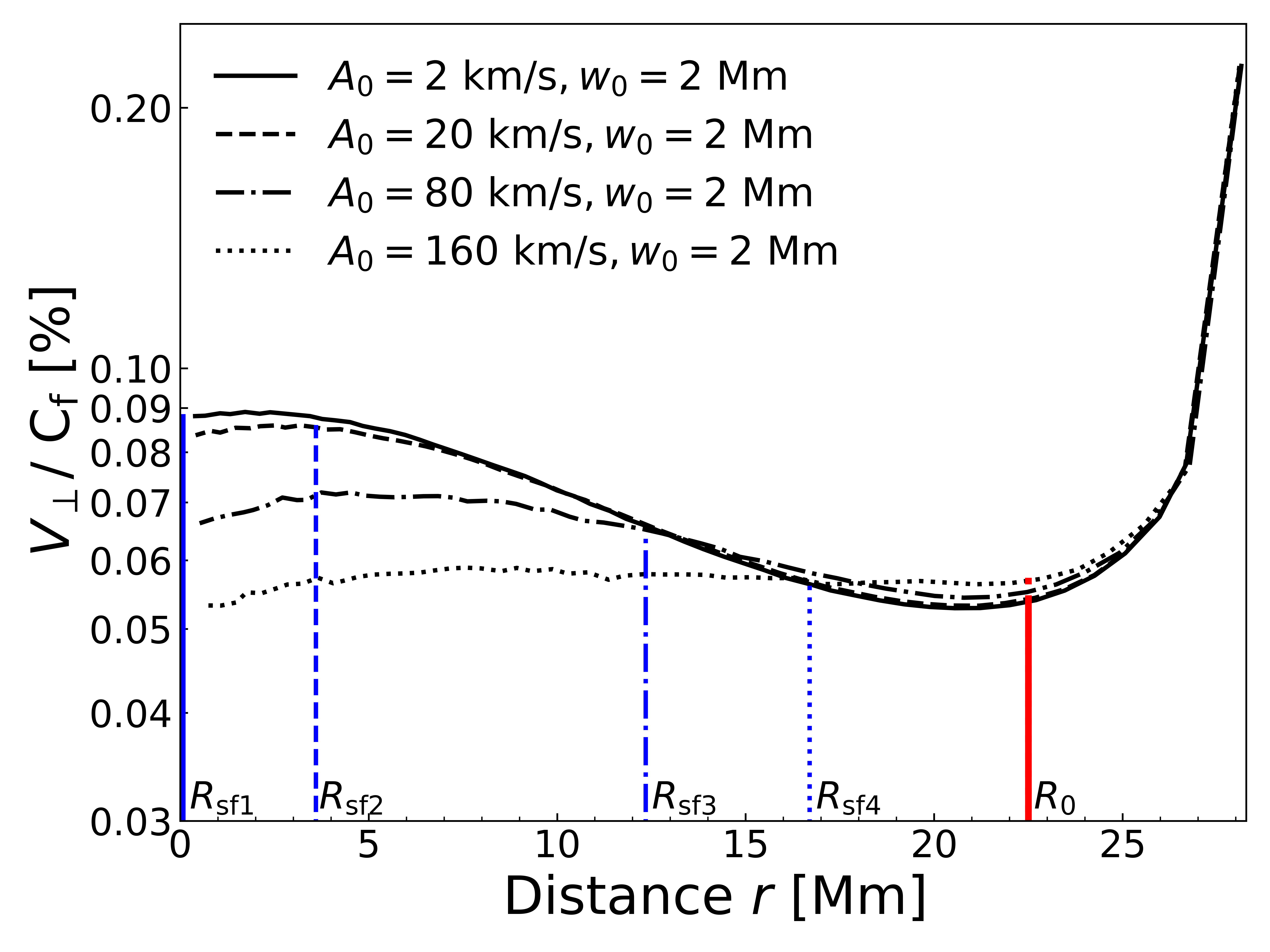}
    \includegraphics[width=0.49\linewidth]{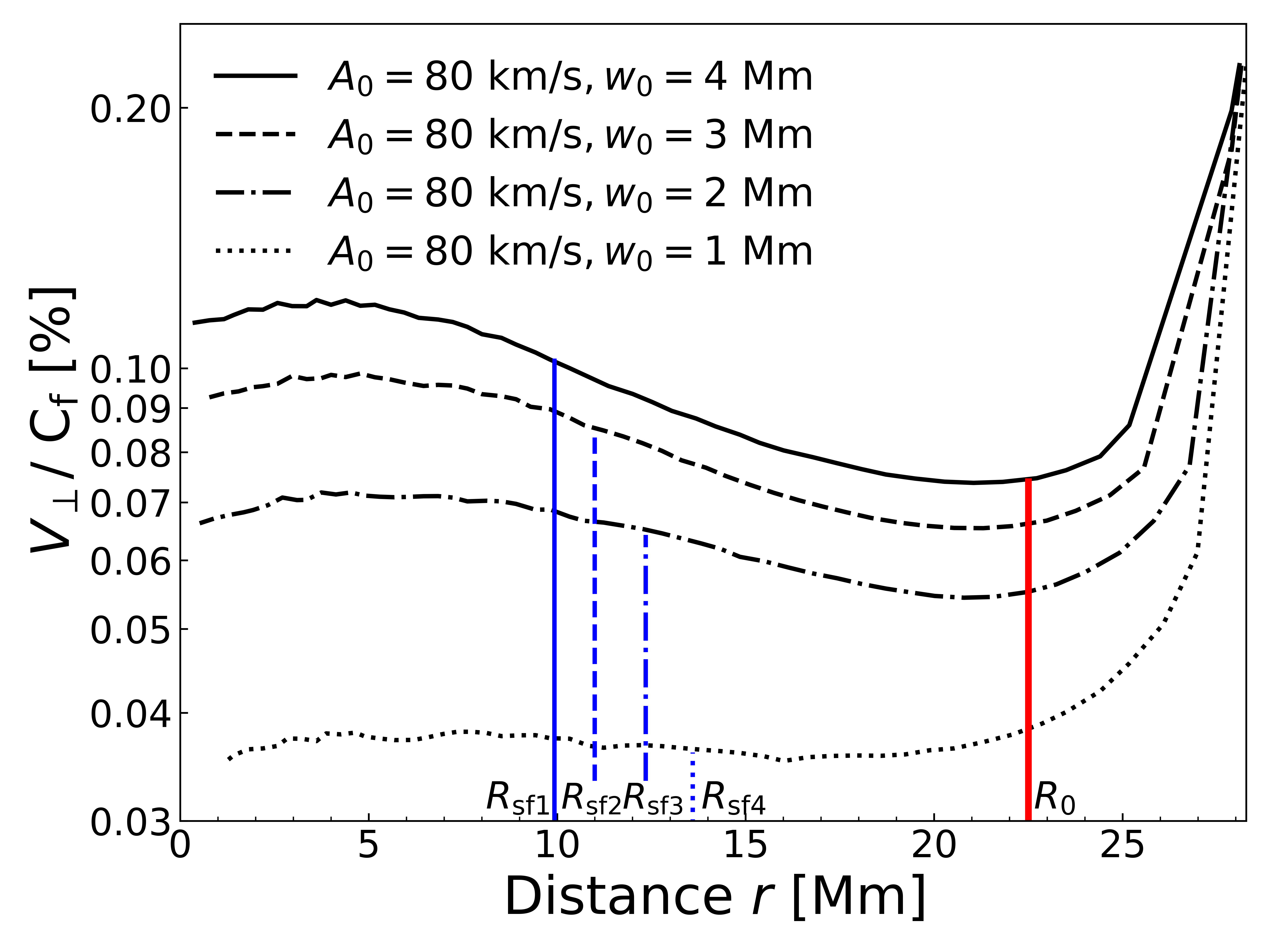}
    \caption{The dependence of the relative amplitude of a fast wave pulse as it approaches a magnetic null point along a magnetic bisector for different initial amplitudes $A_\mathrm{0}$ (left) and widths $w_\mathrm{0}$ (right). Each curve is normalised to its initial amplitude. The vertical axis is plotted on a logarithmic scale. The red and blue vertical lines mark the initial position $R_\mathrm{0}$ and the predicted steepening distance $R_\mathrm{sf}$ from the analytical solution, respectively.}
    \label{fig:vperp2}
\end{figure}

\added{We performed a series of parametric experiments to study the shock formation distance of fast magnetoacoustic wave.} 
Figure~\ref{fig:vperp2} shows the evolution of the pulse amplitude along the magnetic bisector as a function of the distance from the null point. Near the initial location, the amplitude decreases rapidly as a result of the cylindrical expansion. At a certain distance, the segment of the wave front propagating along the bisector becomes planar because of the refraction caused by the equilibrium non-uniformity. Thus, the pulse decrease\added{s} in the relative amplitude then transitions into an increase, up to the point of shock formation. Beyond this point, the wave amplitude undergoes nonlinear damping. As discussed in Section~3\ref{sec:nl}, the location to the shock formation is determined by the initial amplitude $A_0$ and the spatial wave number. The latter parameter is inversely proportional to the effective wavelength, i.e. the initial pulse width $w_0$. 

To estimate the steepening distance via the analytic solution Eq.~(\ref{Eq:chars4}), we choose $r=22.5$~Mm as the starting point $R_0$, where geometric expansion is no longer dominant.  We then extract the waveform parameters ($A_0,\ w_0$) at the time the pulse peak reaches $R_0$ from the numerical simulations. The predicted steepening distances $R_\mathrm{sf}$ are indicated by the blue vertical lines in Figure~\ref{fig:vperp2}. The theoretical values show good agreement with the numerical relative velocity shape. For waves with very small initial velocity, the predicted  steepening  occurs only very close to the null ($R_\mathrm{sf} \approx 0$ ).
While waves with larger initial amplitude and smaller width steepen into shocks at obviously larger distances from the null point.

\begin{figure}
    \centering
    \includegraphics[width=0.49\linewidth]{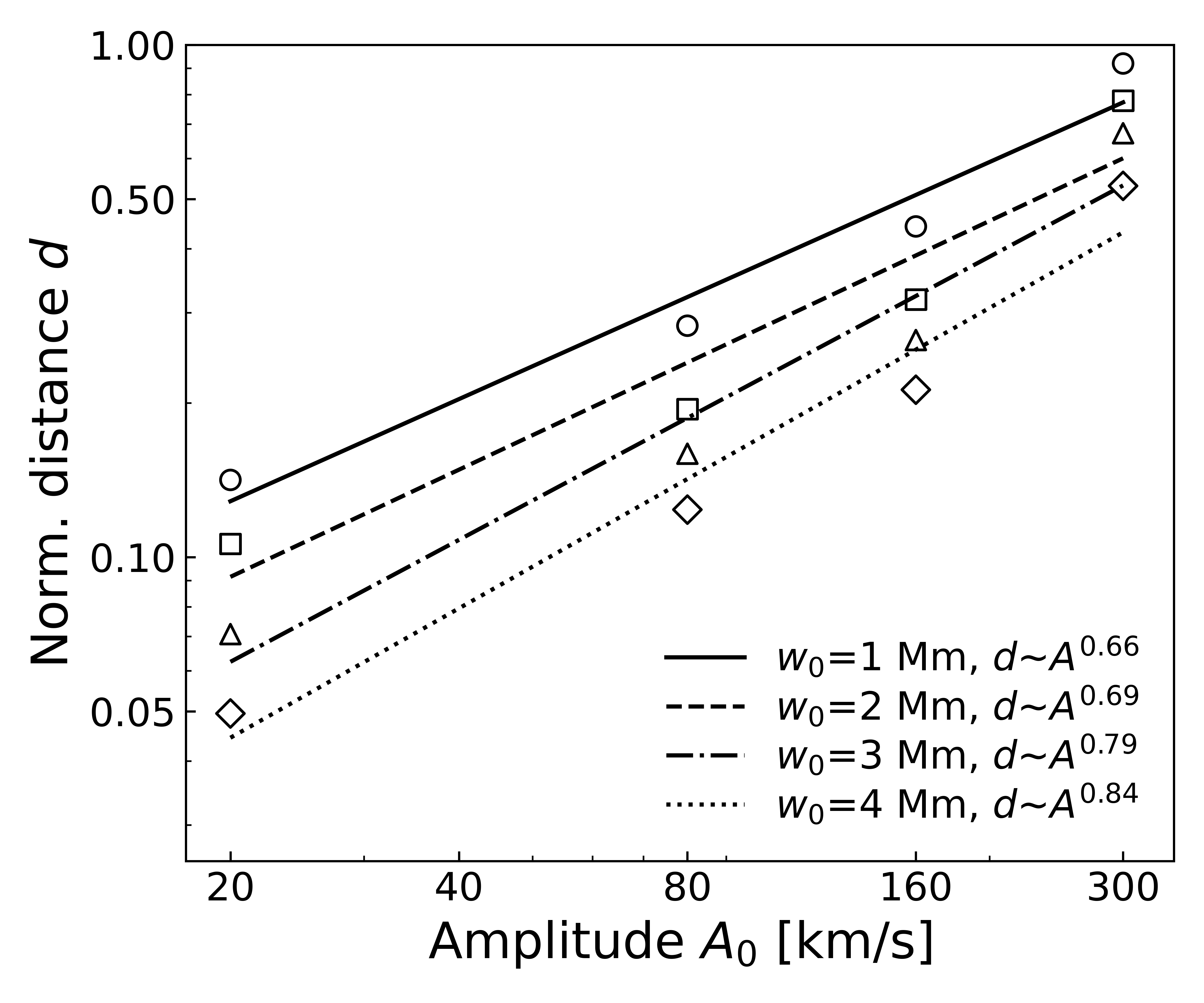}
    \includegraphics[width=0.49\linewidth]{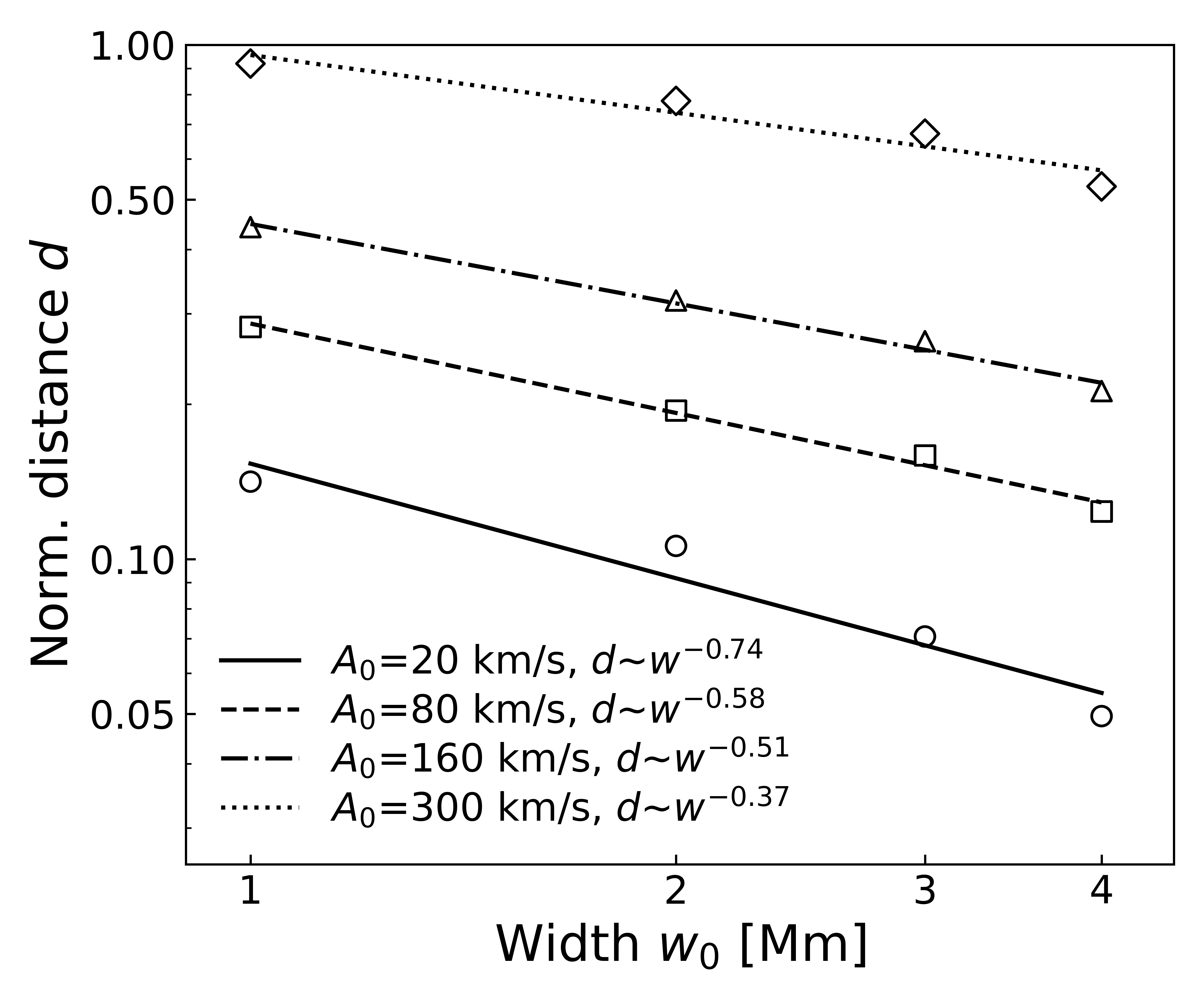}
    \caption{Normalised steepening distance $d$ from magnetic null point as a function of the initial amplitudes $A_0$ (left) and widths $w_0$ (right) in log-log plots. The slanted lines indicate the power-law fits obtained for different $A_0$ and $w_0$ combinations.}
    \label{fig:denp_AW}
\end{figure}

\added{Adopting the same criterion as } \cite{2011A&A...531A..63G}, \added{we also estimated the empirical dependence of the shock formation distance on the amplitude and the width of the initial pulse from numerical simulations. The shocks were assumed to form when the relative amplitude of the pulse began to decrease due to nonlinear damping. Figure \ref{fig:denp_AW} demonstrates the power-law fits of the steepening distance for different combinations of the initial amplitude and pulse width. Applying the least-squares approximation method and performing a joint fit, our parametric analysis yields $d \sim A^{0.75 \pm 0.05}/w ^{0.55 \pm 0.10}$. Notably, the amplitude dependence is more than twice as strong as that reported by} \cite{2011A&A...531A..63G}. 

\section{Conclusions and discussion}
\label{sec:con}
We modelled the interaction of a nonlinear fast wave pulse with a magnetic null point without the guiding field. The pulse is excited at a point located outside the $\beta=1$ distance. As expected, the fast wave pulse with an initially circular wave \added{front} is subject to refraction caused by the non-uniformity of the fast speed, turning the wave front towards the null point. The segment of the external fast wave front, which approaches the null point, is almost planar, which indicates the need to reconsider the results obtained for a circularly symmetric wave front, obtained in \cite{2011A&A...531A..63G}.

The planar shape of the wave front segment that approaches the null point along the magnetic bisector, i.e., across the local equilibrium magnetic field, allows us to employ a simple 1D model for qualitative estimates of the pulse evolution. It is shown that the fast wave pulse becomes narrower and its relative amplitude defined as the ratio of the plasma velocity to the local value of the fast speed, increases. The increase in the wave amplitude makes nonlinear effects important. The fast wave pulse undergoes nonlinear steepening and the following nonlinear dissipation. Waves with higher amplitudes and shorter wavelengths experience shock formation at a larger distance from the null point, i.e., earlier along the inward path. Results of 2D MHD numerical simulations are consistent with the qualitative 1D estimates. 
Furthermore, fast wave pulses are accompanied by perturbations of the electric current density, which reach\added{es} highest amplitudes in the vicinity of shocks. This picture is consistent with the results obtained in \cite{2011A&A...531A..63G} for incoming fast waves with circular (cylindrical) wave fronts. \added{However, the empirical dependence of the steepening distance obtained in our study under the planar-wave regime, $d \sim A^{0.75}/ w ^{0.55}$, shows a stronger sensitivity to the initial amplitude. Based on this, planar wave fronts are expected to steepen and dissipate nonlinearly at earlier stages than circular ones, implying that in the latter case, current enhancements would appear closer to the null point.}

As suggested by \cite{2006A&A...452..343N, 2011A&A...531A..63G}, an incoming fast wave excited by a flare can trigger magnetic reconnection at a remote magnetic null point, and thus produce a secondary (\lq\lq daughter\rq\rq) flare, i.e. cause the phenomenon of sympathetic flares. Our results indicate that this pathway is non-trivial: larger-amplitude incoming waves do not necessarily reach the null because nonlinear steepening rapidly forms shocks, followed by strong nonlinear (shock) dissipation. For successful triggering of a daughter flare, the fast wave must develop into a shock in the vicinity of the null, which requires a specific combination of the amplitude and wave number. This may help explain the relative rarity of sympathetic flares.

\added{In the context of the effect under consideration, the key parameters are the wave amplitude and wavelength near the null point. These are determined by the amplitude of the driver and its distance from the null point. Additionally, stratification and magnetic field geometry near the driver can lead to wave refraction, which also influences the amplitude of the incident wave. This makes a direct comparison between the initial amplitude and observational values difficult. Nevertheless, the initial amplitudes used in our study, 20--160~\kms, corresponding to 4.8--38.5\% of the local Alfv\'en speed, are consistent with typical amplitudes of coronal waves} (e.g., \cite{2012SoPh..281..187P}).

Obviously, the approximate two-dimensional model used here captures only the basic features of this process. A deeper understanding will require fully 3D equilibria (e.g., spine–fan topology) and 3D wave propagation. 
\added{The effects missing in our modelling are the fast wave refraction or focussing in the ignorable dimension, and coupling with Alfv\'en waves at resonant layers. Nevertheless, results of our 2D modelling should remain relevant as the local approximation in the vicinity of the null point or line.}
Furthermore, a possible relevance of this mechanism to sympathetic eruptions is also an interesting question. 

An interesting by-product of this study is the emergence of two successive spikes in the electric current density within an impulsively excited fast magnetoacoustic pulse, \added{see Figure~\ref{fig:vperp}. The leading and trailing current spikes correspond to the compression and rarefaction shocks in the incident fast wave pulse.} This two-spike structure could imprint a double-peaked profile on non-thermal flare emission via two closely spaced episodes of enhanced dissipation/particle acceleration, and may therefore help explain the double-peaked quasiperiodic pulsations reported in some events (e.g., \cite{1987ApJ...321.1031T}). A quantitative test would require forward modelling of the emission and transport.

\ack{
The research was sponsored by the DynaSun project under the Horizon Europe programme of the European Union under grant agreement (no. 101131534). Views and opinions expressed are however those of the authors only and do not necessarily reflect those of the European Union and therefore the European Union cannot be held responsible for them. 
Y.Z. and V.M.N. acknowledge funding from UK Research and Innovation under the UK government the Horizon Europe funding guarantee EP/Y037456/1. M.C. and A.C.  acknowledge support from SECyT (UNC) under grant number
33820230100116CB.
}
%%%%%%%%%% Insert bibliography here %%%%%%%%%%%%%%

\bibliographystyle{RS}
\bibliography{shocks_on_x_point}

\listofchanges
\end{document}